\documentclass[prd,showpacs,preprintnumbers,amsmath,amssymb,superscriptaddress]{revtex4}
\usepackage{graphics}
\usepackage{rotating}

\newcommand{\lpert}{\delta_{\rm L}}
\newcommand{\epert}{\delta_{\rm E}}

\begin{document}

\title{Cosmological Perturbations in Elastic Dark Energy Models}
\author{Richard A. Battye}
\affiliation{Jodrell Bank Observatory, Department of Physics and
Astronomy, University of Manchester, Macclesfield, Cheshire SK11
9DL, UK}
\author{Adam Moss}
\affiliation{Jodrell Bank Observatory, Department of Physics and
Astronomy, University of Manchester, Macclesfield, Cheshire SK11
9DL, UK}
\affiliation{School of Physics and
Astronomy, University of British Columbia, Vancouver, British Columbia, V6T 1Z1,
 Canada}
\date{28/3/2007}
\preprint{}
\pacs{}

\begin{abstract}
We discuss the general framework for a perfect continuum medium in cosmology and show that an interesting generalization of the fluids normally used is for the medium to have rigidity and, hence, be analogous to an elastic solid. Such models can provide perfect, adiabatic fluids which are stable even when the pressure is negative, if the rigidity is sufficiently large, making them natural candidates to describe the dark energy. In fact, if the medium is adiabatic and isotropic, they provide the most general description of linearized perturbations. We derive the equations of motion and wave propagation speeds in the isotropic case. We point out that anisotropic models can also be incorporated within the same formalism and that they are classified by the standard Bravais Lattices. We identify the adiabatic and isocurvature modes allowed in both the scalar and vector sectors and discuss the predictions they make for CMB and matter power spectra. We comment on the relationship between these models and other fluid-based approaches to dark energy, and discuss a possible microphysical manifestation of this class of models as a continuum description of defect-dominated scenarios.
\end{abstract}

\maketitle

\section{Introduction}
\renewcommand{\arraystretch}{1.6}

There are four stress-energy components in the presently preferred standard cosmological model~\cite{Spergel:2006hy,Hinshaw:2006ia,Page:2006hz}: the baryons which constitute the visible universe, radiation (photons and neutrinos), cold dark matter (CDM) and some unknown component, often called dark energy, which gives rise to the cosmic acceleration~\cite{Perlmutter:1996ds,Riess:1998cb,Perlmutter:1998np,Riess:2001gk,Astier:2005qq}. Within the codes~\cite{Seljak:1996is,Lewis:1999bs} used to make predictions for specific models, the CDM is modelled as a pressureless, perfect fluid and the radiation components are modelled as black-body radiation gases. However, many of the qualitative features of the observed power spectra, for example, the acoustic peaks in the angular power spectrum of the cosmic microwave background (CMB), can be derived by ignoring the higher order moments of the photon and neutrino distributions and treating them as a perfect fluid with $P=\rho/3$, where $P$ is the pressure and $\rho$ is the density. Hence, most of the important components of the universe are well approximated by perfect fluids.

The microphysical origin of the cosmic acceleration is still a mystery. The simplest, and probably most popular, explanation is a cosmological constant, although this has some well documented fine-tuning problems~\cite{Carroll:1991mt,Carroll:2000fy}. Various ideas exist, collectively known as dark energy (for a recent review of dark energy models, see ref.~\cite{Copeland:2006wr}), whereby the additional stress-energy component is provided by a slowly rolling scalar field~\cite{Ratra:1987rm,Wetterich:1987fm,Caldwell:1997ii}, known as as Quintessence, or a lattice of topological defects~\cite{Vilenkin:1984rt,Kibble:1985tf,Bucher:1998mh,Battye:1999eq}. Alternative explanations require the modification of gravity by the inclusion of non-minimal coupling between matter and gravity~\cite{Amendola:1999er}, extra-dimensions (for example, ref.~\cite{Dvali:2000hr}) and modifications to the Einstein-Hilbert action~\cite{Capozziello:2003tk,Carroll:2003wy,Carroll:2004de}.

In this paper we shall consider a generalization of the perfect continuum fluid approach to describe dark energy. We have already pointed out that perfect fluids can be used to describe the essential properties of the radiation and CDM in the universe, and it seems reasonable to consider the possibility that the dark energy can also be understood in the same way. This will require us to go back to the fundamentals of how formulate a generalized continuum medium in General Relativity and derive equations for the perturbations in the medium. These are essential in computing accurate predictions for observed power spectra~\cite{Caldwell:1997ii,Battye:1999eq,Weller:2003hw,Bean:2003fb}. We will show that the most obvious generalization of a perfect fluid is to allow for rigidity of the medium in a way analogous to a continuum elastic solid~\cite{Bucher:1998mh}, and that this can be stable if the rigidity is sufficiently large even if the pressure is negative. For most cosmological observations (for example, the CMB or large-scale structure measurements) we only need to consider the linearized regime. We will argue that the generalization of fluids to include rigidity, is the most general possibility for an adiabatic medium at linearized order.
 
In many ways the concept of dark energy is similar to the idea of the Aether postulated in the late 19th century: there is something about the laws of physics which appears awry and we postulate a medium to try and solve the problem~\footnote{In fact, as we will discuss in the final section the elastic medium approach is more than just qualitatively similar to the idea of the Aether. We hope, of course, that it is more successful at explaining observational facts than the Aether.}. The approach we propose is to formulate generalized properties of such a medium, define ways of computing power spectra and then compare them with observations (see ref.~\cite{Battye:2005mm} for the status of this kind of model after the first year WMAP data).

The idea of generalized dark matter/energy has been considered previously by a number of authors~\cite{Hu:1998kj,Hu:1998tj,Bucher:1998mh,Battye:1999eq,Weller:2003hw,Bean:2003fb,Koivisto:2005mm}. We will attempt to discuss how the approaches suggested by others relate to those presented here. The broad difference between our approach and those taken previously is that we will derive the equations describing perturbations in the medium from a set of well defined physical assumptions. These equations are closed and there is no freedom, except the strength of the rigidity, to play with.

The original derivation of the equations presented here~\cite{Bucher:1998mh}, was motivated by the idea that the dark energy could be a lattice of topological defects (cosmic strings or domain walls) formed at a low energy phase transition. We will explain how such a lattice can provide a possible microphysical model for an elastic dark energy model. However, we believe that the formulation of the elastic dark energy models is more general and should not be thought of as being necessarily linked to these specific type of models which predict very specific values for $w=P/\rho$. In particular, the elastic dark energy models include CDM and a cosmological constant as limiting cases allowing them to provide an interesting phenomenology.

The paper is organized as follows. In section~\ref{sec:dynamics} we discuss the formulation of a generalized medium and derive the equations of motion. We then go on to identify the various adiabatic and isocurvature modes in section~\ref{sec:sdemodes} and present the cosmological signatures expected in the CMB and matter power spectra in section~\ref{sec:cosmo}. In section~\ref{sec:micro} we explain the basic ideas of defect-dominated scenarios for dark energy and explain how they relate to the elastic dark energy models.

\section{Cosmological Dynamics of a Generalized Medium} \label{sec:dynamics}

In Section~\ref{sec:materialspace} we review the medium representation concept and the scheme for specifying perturbations in a general medium. In~\ref{sec:cosmoeqns} we apply this formalism to compute the cosmological equations of motion for an isotropic elastic medium and in~\ref{sec:evalsound} evaluate the propagation speeds of perturbations. We then decompose the perturbations in terms of harmonic basis functions in~\ref{sec:harmdecomp} and~\ref{sec:svt_split} to give the full set of Einstein and energy-momentum conservation equations. In Section~\ref{sec:genfluid} we compare our results to other treatments of  generalized fluids, and in~\ref{sec:aniso} discuss briefly how the formalism can be applied to anisotropic perturbations.   

\subsection{Medium Representation Concept} \label{sec:materialspace}

This section provides an overview of work on relativistic elastic media in the context of neutrons stars  by Carter and others~\cite{Carter:1972cq,Carter:1977qf,Carter:1980c,Carter:1982xm}, and details how this  elegant formalism to describe the mechanics of a generalized medium can be applied to cosmology. In Newtonian theory it is natural to define the properties of a medium, such as the density and pressure, in terms of coordinates of a three-dimensional Euclidean space at some given instant of time. In General Relativity, however, there is no general time-slicing of the four-dimensional space-time manifold $\mathcal{M}$ which can be used to specify the material state. Therefore, it is necessary to consider the projection $\mathcal{P}$: $\mathcal{M}  \rightarrow \mathcal{H}$ of $\mathcal{M}$  onto a three-dimensional manifold $\mathcal{H}$ whose elements represent particles of the medium~\cite{Carter:1972cq}. The inverse image $\mathcal{P}^{-1}(X) \subset \mathcal{M}$ of a point $X \in \mathcal{H}$ can then interpreted as the worldline of the particle represented by $X$. The inverse image determines a one-to-one mapping between medium tensors defined on $\mathcal{H}$ and the set of tensors on $\mathcal{M}$ which are orthogonal to the congruence of wordlines. This mapping is important as it allows the intrinsic properties of the material medium to be defined on $\mathcal{H}$, whilst the spacetime evolution is described by tensor fields on $\mathcal{M}$. Moreover, it allows the definition of spacetime tensors in terms the tensors defined on $\mathcal{H}$. We will use Greek indices $\mu, \nu...$ to label tensors on $\mathcal{M}$ and Roman indices $A,B...$ to label tensors on $\mathcal{H}$. 

We define $\gamma_{AB}$ to be the metric on $\mathcal{H}$ which quantifies the strain of the medium. At each point of $\mathcal{H}$, we will assume that there are functions determining the density $\rho$ and pressure $P^{AB}$ which can be expressed in terms of $\gamma_{AB}$ (one of the properties of a perfect medium) and, most probably, related by an equation of state $P^{AB}(\rho)$. Under these assumptions one can show that 
\begin{equation} \label{eqn:materialpressure}
P^{AB} = -2|\gamma|^{-1/2} {\partial \over \partial(\gamma_{AB})} (|\gamma|^{1/2} \rho) \, ,
\end{equation}
where $|\gamma|$  is the determinant of $\gamma_{\rm A B}$, and by taking a second derivative of $\rho$ with respect to $\gamma_{\rm A B}$, one can define
\begin{equation} \label{eqn:materialelasticity}
E^{ABCD} = -2|\gamma|^{-1/2} {\partial \over \partial(\gamma_{AB})} (|\gamma|^{1/2} P^{CD}) \, ,
\end{equation}
the classical elasticity tensor which satisfies $E^{ABCD}=E^{(AB)(CD)}=E^{CDAB}$. We will see that defining these two tensors is sufficient to understand linearized perturbations of the medium.

\begin{figure} 
\centering
\mbox{\resizebox{0.4\textwidth}{!}{\includegraphics{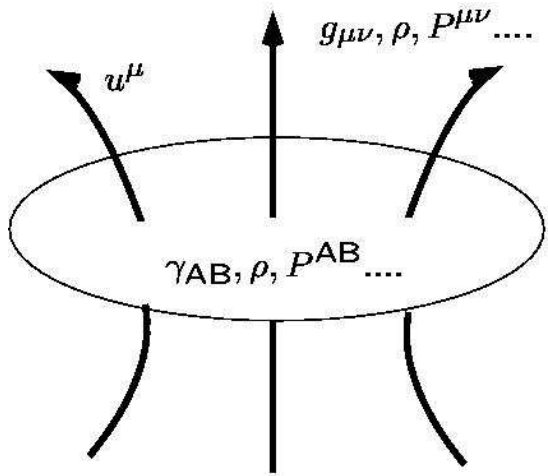}}}
\caption{\label{fig:matflow} Evolution of a material along the flow lines $u^{\mu}$. Physical quantities are defined on the material manifold $\mathcal{H}$ in terms of internal co-ordinates $A, B...$, and there is a direct mapping of spacetime tensors orthogonal to the flow and tensors on $\mathcal{H}$. Parallel components of the spacetime tensors do not contribute to the material projection.}
\end{figure}

In order to make correspondence with the spacetime $\mathcal{M}$, we define $\mathcal{P}_{\mu}^{A}$ to be a bi-tensorial projection operator which projects a tensor in $\mathcal{H}$ into $\mathcal{M}$, or vice-versa, then one can define the projected metric tensor $\gamma_{\mu\nu}$ by
\begin{equation}
\gamma_{\mu\nu}=\mathcal{P}_{\mu}^{A}\mathcal{P}_{\nu}^{B}\gamma_{AB}\,.
\end{equation}
The field of flow vectors $u^{\mu}$ tangent to wordlines in $\mathcal{M}$ are normalized by the condition 
\begin{equation}
u^{\mu} u_{\mu} = -1 \, ,
\end{equation}
and the projection is orthogonal to them, that is, $u^{\mu}\mathcal{P}_{\mu}^A=0$. This allows us to construct $\gamma_{\mu\nu}$ as an orthogonal projection tensor 
\begin{equation}
\gamma_{\mu \nu} = g_{\mu \nu} + u_{\mu} u_{\nu} \, ,
\end{equation}
providing an alternative, but equivalent, definition for $\gamma_{\mu\nu}$. It acts as the  tensor which projects other tensors onto the tangent subspace orthogonal to the wordlines. For example, any vector $v^{\mu}$ can then be decomposed into components orthogonal and parallel to the flow by $v^{\mu}= _\perp\!\!\!v^{\mu}+ v^{\parallel} u^{\mu}$, where $_\perp \! v^{\mu} = \gamma^{\mu}_{\nu} v^{\nu}$ and $  v^{\parallel} = - u_{\mu} v^{\mu}$. This allows physically relevant spacetime tensors to be decomposed into the parallel component, which does not contribute to the material projection, and the orthogonal part, which does. This framework is summarized schematically in Fig.~\ref{fig:matflow}.

One can use the projection tensor to construct the spacetime pressure tensor, $P^{\mu\nu}$, from that on $\mathcal{H}$
\begin{equation}
P_{\mu\nu}={\cal P}_{\mu}^{A}{\cal P}_{\nu}^{B}P_{AB}\,,
\end{equation}
which satisfies $u^{\mu}P_{\mu\nu}=0$. The same can be done for the elasticity tensor $E^{\mu\nu\rho\sigma}$ which satisfies
\begin{equation}
E^{\mu \nu \rho \sigma} = E^{(\mu \nu) (\rho \sigma)} = E^{\rho \sigma \mu \nu}, \hspace{0.5cm} E^{\mu \nu \rho \sigma}u_{\sigma}=0 \, .
\end{equation}

One must also specify how perturbations in $\mathcal{H}$ are related to those in $\mathcal{M}$ in order to deal with dynamics. The convected differential $d[\, \,]$ is an important quantity, as it is the spacetime tensor corresponding to the Lagrangian (wordline preserving) material variation on $\mathcal{H}$. In the case of a vector, there is the bijection $d[v^{\mu}] \longleftrightarrow \{ \lpert \, _\perp \! v^{\rm A}, \, \lpert  v^{\parallel} \}$, where $\lpert$ is the Lagrangian variation. The convected differential can be evaluated in terms of Lagrangian differentials on $\mathcal{M}$ by projecting the orthogonal part of the perturbation. This gives the explicit relationship~\cite{Carter:1980c}
\begin{equation} \label{eqn:convectderiv}
d[T^{\mu...}_{\nu...}] = \lpert T^{\mu...}_{\nu...}+T^{\mu...}_{\rho...}u^{\rho} \lpert u_{\nu} - T^{\rho...}_{\nu...}u^{\mu} \lpert u_{\rho} \, ,
\end{equation}
where $T^{\mu...}_{\nu...}$ is a general mixed tensor.

The next stage is to relate the material variations in terms of spacetime tensors. The most direct way to do this is to use the convected derivative, which relates Lagrangian and material variations. Applying (\ref{eqn:convectderiv}) to (\ref{eqn:materialpressure}) and (\ref{eqn:materialelasticity}) one obtains~\cite{Carter:1982xm}
\begin{equation} \label{eqn:spacetimedensity}
\lpert \, \rho = - \frac{1}{2} ( P^{\mu \nu} + \rho \gamma^{\mu \nu} ) \lpert \, g_{\mu \nu}\, ,
\end{equation} 
\begin{equation} \label{eqn:spacetimepressure}
\lpert P^{\mu \nu} = -\frac{1}{2} (  E^{\mu \nu \rho \sigma} + P^{\mu \nu} \gamma^{\rho \sigma} -  4 P^{\rho ( \mu} u^{\nu)} u^{\sigma}  ) \lpert \, g_{\rho \sigma}\, ,
\end{equation} 
where extra terms are induced in (\ref{eqn:spacetimepressure}) due to (\ref{eqn:convectderiv}).

We can now begin to discuss the dynamics of an elastic medium. A {\it perfect elastic medium} is defined by the condition that the energy-momentum tensor $T^{\mu \nu}$ is a material function of the metric tensor with respect to the flow field~\cite{Carter:1982xm}, meaning that it takes the form
\begin{equation} \label{eqn:perfectem}
T^{\mu \nu} = \rho u^{\mu} u^{\nu} + P^{\mu \nu} \,.
\end{equation}

The variation in the energy-momentum tensor can be obtained by using (\ref{eqn:perfectem}) in conjunction with (\ref{eqn:spacetimedensity}) and (\ref{eqn:spacetimepressure}) to give
\begin{equation} \label{eqn:emperturb}
\lpert T^{\mu \nu} = - \frac{1}{2} ( W^{\mu \nu \rho \sigma} + T^{\mu \nu} g^{\rho \sigma} ) \lpert \, g_{\rho \sigma}\,,
\end{equation}
where the non-orthogonal elasticity tensor $W^{\mu \nu \rho \sigma}$ can be decomposed as~\cite{Friedman:1975fs}
\begin{equation}
W^{\mu \nu \rho \sigma} = E^{\mu \nu \rho \sigma} + P^{\mu \nu} u^{\rho} u^{\sigma} +   P^{\rho \sigma} u^{\mu} u^{\nu} - P^{\mu \rho} u^{\sigma} u^{\nu} - P^{\mu \sigma} u^{\nu} u^{\rho} - P^{\nu \rho} u^{\mu} u^{\sigma} - P^{\nu \sigma} u^{\mu} u^{\rho}- \rho u^{\mu} u^{\nu} u^{\rho} u^{\sigma} \, ,
\end{equation}
and has the same symmetry properties as the ordinary elasticity tensor. This allows the variations to be conveniently written in terms of functional derivatives of the Lagrangian with respect to the metric via
\begin{equation}
T^{\mu\nu}=-2|g|^{-1/2}{\delta \over \lpert g_{\mu\nu}}(|g|^{1/2}{\cal L})\,,
\end{equation}  
\begin{equation} \label{eqn:defnw}
W^{\mu\nu\rho\sigma}=4|g|^{-1/2}{\delta \over \lpert g_{\rho\sigma}}{\delta \over \lpert g_{\mu\nu}}(|g|^{1/2}{\cal L}) =-2|g|^{-1/2}{\delta \over \lpert g_{\rho\sigma}}(|g|^{1/2}T^{\mu\nu})\,,
\end{equation}
where $|g|$ is the determinant of the metric. 

It is often more convenient to describe perturbations being fixed with respect to some background space. If the vector field $\xi^{\mu}$ is the infinitesimal displacement of the wordlines with respect to the fixed background space, then the difference between Lagrangian $\lpert$ and fixed (Eulerian) variations $\epert$ is given by
\begin{equation} \label{eqn:lageul}
\lpert = \epert + \mathcal{L}_{\xi} \, ,
\end{equation}
where $\mathcal{L}_{\xi}$ is the Lie derivative. Such a displacement could be removed by using a mapping of the perturbed space onto the background space, but it will be more convenient to set the gauge by some other means. In the following section, for example, we use the synchronous gauge to derive the cosmological equations of motion.  For the particular case of the metric tensor then the relationship (\ref{eqn:lageul}) gives 
\begin{equation}
\lpert \, g_{\mu \nu} = \epert \, g_{\mu \nu} + 2 \nabla_{(\mu} \xi_{\nu)}\,,
\end{equation}
where $\epert \, g_{\mu \nu}$ is the Eulerian variation of the metric tensor.

The equation of motion for the vector field $\xi^{\mu}$ gives the complete system of equations for the perturbations. Evaluation of the Lagrangian variation  $\lpert ({\gamma^{\mu}}_{\nu} \nabla_{\mu} T^{\mu \nu})$ gives
\begin{eqnarray} \label{eqn:eom}
({{A^{\mu(\nu}}}{_{\rho}{^{\sigma)}}} - (\rho {\gamma^{\mu}}_{\rho} + {P^{\mu}}_{\rho}) u^{\nu} u^{\sigma}) \lpert {\Gamma^{\rho}}_{\nu \sigma} + \frac{1}{2} {\gamma^{\mu}}_{\rho}  {\gamma^{\alpha}}_{\nu}  {\gamma^{\beta}}_{\sigma} (\lpert \, g_{\alpha \beta}) \nabla_{\tau} E^{\rho \tau \nu \sigma} = \\ \nonumber
(P^{\mu \nu} \dot{u}^{\sigma} - \frac{1}{2} P^{\nu \sigma} \dot{u}^{\mu} - 2{{A^{\mu(\nu}}}{_{\rho}{^{\tau)}}} {v^{\rho}}_{\tau}  u^{\sigma} +  (\rho {\gamma^{\mu}}_{\rho} + {P^{\mu}}_{\rho}) \dot{u}^{\rho} u^{\nu} u^{\sigma} ) \lpert \, g_{\nu \sigma}\,, 
\end{eqnarray}
where dots denote covariant differentiation with respect to the flow (that is, $u^{\mu} \nabla_{\mu}$) and
\begin{equation} \label{eqn:hadamardtens}
{{A^{\mu \nu}}}{_{\rho}{^{\sigma }}} = {{E^{\mu \nu}}}{_{\rho}{^{\sigma}}} -  {\gamma^{\mu}}_{\rho} P^{\nu \sigma}\,,
\end{equation}
is the relativistic Hadamard elasticity tensor which forms the characteristic equation needed to evaluate the sound speeds in the medium~\cite{Carter:1973}.  This tensor obeys the symmetry and orthogonality conditions
\begin{equation}
A^{\mu \nu \rho \sigma} = A^{\rho \sigma \mu \nu}, \hspace{0.5cm} A^{\mu \nu \rho \sigma}u_{\sigma}=0\,.
\end{equation}
Eqn.~(\ref{eqn:eom}) takes the form of a wave equation for the displacement vector $\xi^{\mu}$. Since all components are orthogonal the the flow then the additional degree of freedom in  $\xi^{\mu}$ can be removed by imposing the orthogonality requirement $\xi^{\mu} u_{\mu}=0$. The remaining quantities required to evaluate (\ref{eqn:eom}) are the Lagrangian variation of the connection coefficients, given by
\begin{equation}
\lpert {\Gamma^{\mu}}_{\nu \sigma} = \nabla_{(\nu} \lpert \,{g^{\mu}}_{\sigma)} -\frac{1}{2} \nabla^{\mu} \lpert \, g_{\nu \sigma}\,,
\end{equation}
and the flow gradient tensor given by
\begin{equation}
v_{\mu \nu} = \nabla_{\nu} u_{\mu} + \dot{u}_{\mu} u_{\nu}\,.
\end{equation}

\subsection{Cosmological Equations of Motion for an Isotropic Medium} \label{sec:cosmoeqns}

The perturbed space-time metric takes the form
\begin{equation}
g_{\mu \nu} = a^{2} (\tau) \left[ \eta_{\mu \nu} + h_{\mu \nu} \right], \hspace{0.5cm} h^{\mu \nu} = \eta^{\alpha \mu} \eta^{\beta \nu} h_{\alpha \beta}\,,
\end{equation}
and so the Eulerian component of the metric perturbation is given by $\epert \, g_{\mu \nu} = a^{2} h_{\mu \nu}$. We make use of the synchronous gauge conditions ($h_{00}=h_{0i}=0$) to remove the remaining degree of gauge freedom in the Einstein field equations. At zeroth order the flow vector $u^{\mu}=a^{-1}(1,0,0,0)$, and so the non-zero components of the displacement vector $\xi^{\mu}$ are confined to the spatial part by the orthogonality requirement $\xi^{\mu} u_{\mu}=0$. 

In the synchronous gauge the non-zero components of the Lagrangian variation of the connection are given by
\begin{eqnarray}
\lpert  {\Gamma^{i}}_{00} = \ddot{\xi}^{i} + {\cal H} \dot{\xi}^{i}\,,\quad
\lpert {\Gamma^{0}}_{0i} = {\cal H} \dot{\xi}_{i}\,,\quad
\lpert {\Gamma^{0}}_{ij} = 2 {\cal H} \partial_{(i} \xi_{j)} +  {\cal H} h_{ij} + \dot{h}_{ij}\,,\\ \nonumber
\lpert {\Gamma^{i}}_{0j} = \partial_{j} \dot{\xi}^{i} + \frac{1}{2} {{\dot{h}}^{i}}_{\,j} \,,\quad
\lpert {\Gamma^{i}}_{jk} = \partial_{j} \partial_{k} \xi^{i} -  \delta_{jk} {\cal H} \dot{\xi^{i}} + \partial_{(j}{h^{i}}_{k)} - \frac{1}{2} \partial^{i} h_{jk}\,,
\end{eqnarray}
where dots are now understood to denote derivatives with respect to the conformal time and ${\cal H}$ is the conformal time Hubble parameter. We can now obtain the equations of motion and perturbed energy-momentum sources for the cosmological fluids by inserting the appropriate pressure and elasticity tensor expressions for each component into (\ref{eqn:emperturb}) and (\ref{eqn:eom}).

\subsubsection{Isotropic Perfect Fluid}

In an isotropic perfect fluid the pressure tensor is isotropic and is given in terms of the pressure scalar $P$ by
\begin{equation} \label{eqn:presspf}
P^{\mu \nu} = P \gamma^{\mu \nu}\,,
\end{equation}
while the elasticity tensor is given in terms of the bulk modulus $\beta$ by~\cite{Carter:1972cq}
\begin{equation} \label{eqn:elpf}
E^{\mu \nu \rho \sigma} = (\beta - P) \gamma^{\mu \nu} \gamma^{\rho \sigma} + 2 P \gamma^{\mu ( \rho} \gamma^{\sigma) \nu}\,,
\end{equation}
and the bulk modulus is defined by
\begin{equation}
\beta= (\rho + P ) \frac{d P}{d \rho}\,.
\end{equation}
Substituting these expressions into (\ref{eqn:emperturb}), we obtain the components of the perturbed energy-momentum tensor
\begin{subequations}
\begin{eqnarray}
\epert \, {T^{0}}_{0} &=&  (\rho  + P) \left( \partial_{i} \xi^{i} +  \frac{1}{2} h \right)\,, \\
\epert \, {T^{i}}_{0} &=& -(\rho + P) \dot{\xi}^{i}\,, \\
\epert \, {T^{i}}_{j} &=& - \beta {\delta^{i}}_{j} \left( \partial_{k} \xi^{k} + \frac{1}{2} h \right)\,,
\end{eqnarray}
\end{subequations}
where $h$ is the trace of the metric perturbation. From now onward we do not make the explicit distinction between Eulerian (fixed) and Lagrangian perturbations, as all relevant quantities are evaluated with respect to the fixed coordinates. Substitution of (\ref{eqn:presspf}) and (\ref{eqn:elpf}) into (\ref{eqn:eom}) gives the equation of motion for the displacement vector $\xi^{i}$ as
\begin{equation}
(\rho + P)( \ddot{\xi}^{i} + {\cal H} \dot{\xi}^{i}) - 3 \beta {\cal H} \dot{\xi}^{i} - \beta (\partial^{i} \partial_{j} \xi^{j} +  \partial^{i} h/2)=0\,.
\end{equation}

\subsubsection{Isotropic Perfect Elastic Medium}

We have pointed out that perturbations in an elastic medium can be specified by the pressure tensor $P^{\mu\nu}$ and the elasticity tensor $E^{\mu\nu\rho\sigma}$. In the case of isotropy the pressure tensor is also given by (\ref{eqn:presspf}) since $\gamma^{\mu\nu}$ is the only isotropic tensor of rank two, upto a scaling. The case of rank four, required to describe the elasticity tensor, is more complicated; it is given by 
\begin{equation}
E^{\mu\nu\rho\sigma}=A\gamma^{\mu\nu}\gamma^{\rho\sigma}+B\gamma^{\mu(\rho}\gamma^{\sigma)\nu}\,,
\end{equation}
where $A$ and $B$ are arbitrary parameters. In order to fit in with the definitions used in the previous section, we will define an additional shear contribution by 
\begin{equation} \label{eqn:eliso}
E^{\mu \nu \rho \sigma} = \Sigma^{\mu \nu \rho \sigma} + (\beta - P) \gamma^{\mu \nu} \gamma^{\rho \sigma} + 2 P \gamma^{\mu ( \rho} \gamma^{\sigma) \nu}\,,
\end{equation}
where the shear tensor obeys the symmetry and orthogonality conditions
\begin{equation}
\Sigma^{\mu \nu \rho \sigma} = \Sigma^{(\mu \nu) (\rho \sigma)} = \Sigma^{\rho \sigma \mu \nu}, \hspace{0.5cm} \Sigma^{\mu \nu \rho \sigma}u_{\sigma}=0 \, .
\end{equation}
In an isotropic elastic fluid the shear tensor in terms of a single shear moduli $\mu$ by 
\begin{equation}
\Sigma^{\mu \nu \rho \sigma} = 2 \mu \left(\gamma^{\mu ( \rho} \gamma^{\sigma ) \nu} - \frac{1}{3} \gamma^{\mu \nu} \gamma^{\rho \sigma} \right)\,.
\end{equation}
The contribution of the shear tensor to the elasticity tensor is zero in the perfect fluid case ($\mu = 0$). We again substitute these expressions into (\ref{eqn:emperturb}) to obtain the components of the perturbed energy-momentum tensor
\begin{subequations}
\begin{eqnarray}
{\delta T^{0}}_{0} &=&  (\rho + P) \left( \partial_{i} \xi^{i} + \frac{1}{2} h \right)\,, \label{eqn:T00elast} \\
{\delta T^{i}}_{0} &=& -(\rho + P) \dot{\xi}^{i}\,, \label{eqn:Ti0elast}  \\
{\delta T^{i}}_{j} &=& - {\delta^{i}}_{j} ( \beta - \frac{2}{3} \mu ) \left( \partial_{k} \xi^{k} + \frac{1}{2} h \right) - \mu (2 \partial_{(j} \xi^{i)} + h^{i}_{j})\,. \label{eqn:Tijelast} 
\end{eqnarray}
\end{subequations}
The equation of motion for the displacement vector $\xi^{i}$ is then given by
\begin{equation} \label{eqn:solideom}
(\rho + P)( \ddot{\xi}^{i} + {\cal H} \dot{\xi}^{i}) - 3 \beta  {\cal H} \dot{\xi}^{i} - \beta (\partial^{i} \partial_{j} \xi^{j} + \partial^{i} h/2)-\mu ( \partial^{j} \partial_{j} \xi^{i} +  \partial^{i} \partial_{j} \xi^{j}/3 + \partial^{j} h^{i}_{j} - \partial^{i} h/3) =0\,.
\end{equation}
where (\ref{eqn:T00elast}, \ref{eqn:Ti0elast}, \ref{eqn:Tijelast}) and (\ref{eqn:solideom}) agree with the equations for an elastic medium defined in ref.~\cite{Bucher:1998mh}. If the medium forms at some finite time then boundary conditions imply that $h_{ij} \rightarrow h_{ij} - h_{ij}^{I}$, where $I$ refers to quantities defined at the time of formation of the medium. Since only the second-order variation of the Lagrangian is relevant for linearized perturbations we have shown that the pressure tensor, $P^{\mu \nu}$, and elasticity tensor, $E^{\mu \nu \rho \sigma}$, specify the most general parameterization of perturbations in $T^{\mu \nu}$ under the assumptions discussed earlier.

\subsection{Evaluation of Sound Speeds} \label{sec:evalsound}

This section provides a brief summary of how to compute the propagation speed and polarization directions of sound waves (small perturbations) in a perfectly elastic medium,  which was originally developed in ref.~\cite{Carter:1973}. A sound wave front is defined as the hypersurface across which the acceleration vector $u^{\nu} \nabla_{\nu} u^{\mu}$ has a jump discontinuity (that is, at some coordinate point $x^{\mu}_{0}$ the acceleration is a well defined function in the limits $x^{\mu}_{-}$ and $x^{\mu}_{+}$, but is not defined at $x^{\mu}_{0}$). The flow vector $u^{\mu}$, the functions of state $\rho$, $P^{\mu \nu}$, $E^{\mu \nu \rho \sigma}$, the space-time tensor $g_{\mu \nu}$ and the projection tensor $\gamma_{\mu \nu}$ are continuous across this hypersurface, but the acceleration induces discontinuities in the first derivatives of $\rho$, $P^{\mu \nu}$, $E^{\mu \nu \rho \sigma}$ and $\gamma_{\mu \nu}$.

At the wave front hypersurface $u^{\nu} \nabla_{\nu} u^{\mu}= \alpha l^{\mu}$, where $\alpha$ is the amplitude of the sound wave and $l^{\mu}$ is the polarization vector of the wave front, which satisfies
\begin{equation}
l^{\mu} l_{\mu} = 1, \hspace{0.5cm} l^{\mu} u_{\mu} = 0\,.
\end{equation}
If one introduces a propagation direction vector $v^{\mu}$ satisfying the same orthonormality conditions as the polarization vector, then a characteristic equation
\begin{equation}
[v^{2} (\rho \gamma^{\mu \nu} + P^{\mu \nu}) - Q^{\mu \nu}]\,l_{\nu} = 0\,,
\end{equation}
can be constructed whose eigenvalues give the squared propagation speeds $v^{2}$ in the direction specified by  $v^{\mu}$ and the eigenvectors are the polarization direction(s). The relativistic Fresnel $Q^{\mu \nu}$ tensor is defined by
\begin{equation} \label{eqn:fresnel}
Q^{\mu \nu} = A^{\mu \rho \nu \sigma} v_{\rho} v_{\sigma}\,,
\end{equation}
where the Hadamard tensor $A^{\mu \rho \nu \sigma}$ is defined by (\ref{eqn:hadamardtens}). The Fresnel tensor satisfies the symmetry and orthogonality conditions
\begin{equation}
Q^{\mu \nu} = Q^{(\mu \nu)}, \hspace{0.5cm} Q^{\mu \nu} u_{\nu} = 0\,.
\end{equation}

\subsubsection{Isotropic Perfect Fluid}

The Fresnel tensor can be obtained by substituting the expression for the pressure tensor (\ref{eqn:presspf}) and elasticity tensor (\ref{eqn:elpf}) into (\ref{eqn:hadamardtens}) and then using (\ref{eqn:fresnel}) to give
\begin{equation}
Q^{\mu \nu} = \beta v^{\mu} v^{\nu}\,,
\end{equation}
so the eigenvalue equation becomes
\begin{equation}
[v^{2} (\rho +P) \gamma^{\mu \nu}  - \beta v^{\mu} v^{\nu} ]l_{\nu} = 0\,.
\end{equation}
This equation has a single solution in which the propagation direction is parallel to the polarization vector $(l_{\mu} = v_{\mu})$ and is given by
\begin{equation}
v^{2} = c_{\rm s}^{2} = \frac{\beta}{\rho + P} = \frac{dP}{d \rho}\,,
\end{equation}
 which corresponds to the longitudinal (scalar) sound speed. 

\subsubsection{Isotropic Perfect Elastic Medium}

\begin{figure} 
\centering
\mbox{\resizebox{0.9\textwidth}{!}{\includegraphics{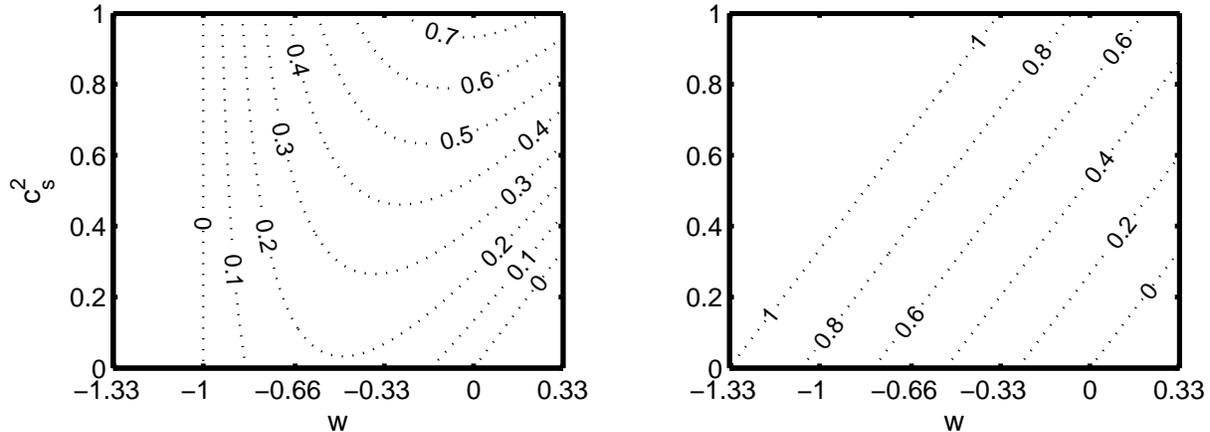}}}
\caption{Allowed parameter space of elastic fluid models in the  $(w,c_{\rm s}^2)$ plane. The left panel shows lines of constant $\mu/\rho$, with the constraint that $\mu/\rho\ge 0$ disallowing regions with $w < -1$ and $w > c_{\rm s}^2$. The right panel shows lines of constant $c_{\rm v}^{2}$, with the constraint that $0 \le c_{\rm v}^{2} \le 1$ disallowing an additional region with $w > c_{\rm s}^2-4/3$.}
\label{fig:sdm}
\end{figure}

In the case of an isotropic elastic medium the Fresnel tensor is given by
\begin{equation}
Q^{\mu \nu} = \left(\beta + \frac{1}{3} \mu\right) v^{\mu} v^{\nu}+ \mu \gamma^{\mu \nu}\,,
\end{equation}
so the eigenvalue equation becomes
\begin{equation}
\left[v^{2} (\rho +P) \gamma^{\mu \nu}  - \mu \gamma^{\mu \nu} - \left(\beta +\frac{1}{3} \mu\right) v^{\mu} v^{\nu} \right]l_{\nu} = 0\,.
\end{equation}
In this case there are two solutions - again there is one where the propagation direction is parallel to the polarization vector --- but also another where the  propagation direction is orthogonal to the polarization vector $(l^{\mu} v_{\mu}=0)$. This additional solution corresponds to a transversely polarized (vector) sound speed. The two solutions are given by
\begin{equation} \label{eqn:soundconst1}
v^2=c_{\rm s}^{2}=\frac{\beta+4 \mu/3}{\rho + P}, \hspace{0.5cm} v^2=c_{\rm v}^{2}=\frac{\mu}{\rho+P}\,, 
\end{equation}
so the two sound speeds are related by
\begin{equation} \label{eqn:soundconst}
c_{\rm s}^{2} = \frac{dP}{d \rho} + \frac{4}{3} c_{\rm v}^{2}\,.
\end{equation}
For an equation of state where $P=w \rho$ it can be seen from (\ref{eqn:soundconst}) that if $\mu/\rho$ is sufficiently large then $c_{\rm s}^2>0$ even if $w$ is negative. This stabilizing property of the shear modulus initially motivated the use of elastic fluids in the framework of dark energy models, where  $w<-1/3$ is required to achieve the observed acceleration~\cite{Bucher:1998mh,Battye:1999eq}. The relationship between the sound speeds and equation of state is shown in Fig.~\ref{fig:sdm}, where we plot lines of constant $\mu/\rho$ and 
$c_{\rm v}^2$ in the $(w,c_{\rm s}^2)$ plane. The constraint that $\mu/\rho \ge 0$ requires that $w \ge -1$, while the constraint that $0 \le c_{v}^{2}\le1$ restricts the allowed value of $c_{\rm s}^2$ for a given $w$. We have assumed that the intrinsic properties of the medium are fixed and have ignored time variations in quantities such as $w$ and $\mu / \rho$. Relaxing this assumption could lead to a well-defined phenomenological model for time-varying dark energy. However, such a model would have to respect the stability conditions discussed above.

\subsection{Harmonic Decomposition of Perturbations} \label{sec:harmdecomp}

Constant time hypersurfaces are homogeneous and isotropic, so it is natural to decompose spatial tensor fields on these hypersurfaces in terms of the irreducible representations of the rotation group SO(3). A spatial tensor field can be decomposed in terms of eigenfunctions of the Laplacian~\cite{Lifshitz:1963ps,Bardeen:1980kt,Abbott:1986ct}
\begin{equation}
\eta^{ij}  Q_{|ij}^{S,V,T} = -k^{2} Q^{S,V,T}\,,
\end{equation}
where the $S, V, T$ index represents irreducible scalar, vector and tensorial quantities and $'|'$ represents covariant differentiation with respect to the three-metric $\eta_{ij}$.  These eigenfunctions form a complete basis in which to expand the tensor field. In flat space, for example, Fourier plane waves provide a local orthonormal basis. Scalars can be constructed from longitudinal type vectors and tensors via~\cite{Hu:1997mn}
\begin{equation}
Q^{S}_{i}=-k^{-1}  Q_{|i}^{S}\,, \hspace{0.5cm} Q^{S}_{ij} = k^{-2} Q_{|ij}^{S} + \frac{1}{3} \eta_{ij} Q^{S}\,,
\end{equation}
and similarly vectors can be constructed from solenoidal type tensors by
\begin{equation}
Q_{ij}^{V} = -(k)^{-1} Q_{(i|j)}^{V}\,, 
\end{equation}
subject to $ Q^{V |i}_{i} = Q^{T |i}_{ij} = Q^{T i}_{i}=0 $. A vector field can, therefore, be decomposed as 
\begin{equation} \label{eqn:vecdecom}
\xi_{i} = \xi^{S} Q^{S}_{i} + \xi^{V} Q_{i}^{V}\,,
\end{equation}
and the general decomposition of a symmetric tensor field is
\begin{equation} \label{eqn:tensdecom}
H_{ij} =  H_{L}^{S} Q^{S} \eta_{ij} +    H_{T}^{S} Q_{ij}^{S} +  H^{V} Q_{ij}^{V}  +  H^{T} Q_{ij}^{T}\,.
\end{equation}
The six degrees of freedom of the tensor field are represented by two scalar parts along with vector and tensor parts each with two degrees of freedom. 

\subsection{Einstein Equations and Conserved Energy-Momentum}  \label{sec:svt_split} 

The perturbed Einstein equations subject to the synchronous gauge conditions are 
\begin{subequations}
\begin{eqnarray}
a^{2} {G^{0}}_{0} &=&   - 3 {\cal H}^2 - {\cal H} \dot{h} + \frac{1}{2} \partial_{i} \partial^{i} h - \frac{1}{2} \partial_{i} \partial_{j} h^{ij} \,, \label{eqn:G00} \\
2 a^{2} {G^{0}}_{i}&=&   \partial_{i}\dot{h}  - \partial_{j}{\dot{h}^{j}}_{\,\,i} \,, \label{eqn:G0i} \\
2 a^{2} {G^{i}}_{0}&=&   \partial_{j}\dot{h}^{ij} - \partial^{i}\dot{h} \,, \label{eqn:Gi0} \\
a^{2} {G^{i}}_{j}&=&    \left( 2 \dot{{\cal H}} - {\cal H}^{2} \right) {\delta^{i}}_{j} +\frac{1}{2} \left( {\ddot{h}^{i}}_{\,\,j} - \ddot{h} {\delta^{i}}_{j} \right) + {\cal H} \left({\dot{h}^{i}}_{\,\,j} - \dot{h} {\delta^{i}}_{j} \right) +\frac{1}{2} \left({\delta^{i}}_{j} \partial_{k} \partial^{k} h  - \partial_{k} \partial^{k} {h^{i}}_{j} \right)   \label{eqn:Gij} \\ \nonumber
 & & + \frac{1}{2} \delta^{ik} \left( \partial_{k} \partial_{l} {h^{l}}_{j} + \partial_{j} \partial_{l} {h^{l}}_{k} - \partial_{k} \partial_{j} h  \right)  - \frac{1}{2} {\delta^{i}}_{j} \partial_{k} \partial_{l} h^{kl}  \,. 
\end{eqnarray}
\end{subequations}
The metric perturbation is parameterized by $h_{ij}= 2 H_{ij}$, where $H_{ij}$ is decomposed according to (\ref{eqn:tensdecom}). In order to compare the elastic fluid  with other fluid based models, we use the parameterization of the the energy-momentum tensor used in ref.~\cite{Hu:1997mn},
\begin{subequations}
\begin{eqnarray}
{T^{0}}_{0} &=& - (\rho+\delta \rho)\,, \label{eqn:comT00} \\
{T^{0}}_{i} &=& (\rho + P) v_{i}\,,\label{eqn:comT0i}  \\
{T^{i}}_{0} &=& -(\rho + P) v^{i}\,, \label{eqn:comTi0} \\
{T^{i}}_{j} &=& (P+\delta P) {\delta^{i}}_{j} + P  {\Pi^{i}}_{j}\,, \label{eqn:comTij}
\end{eqnarray}
\end{subequations}
where $v_{i}$ is the velocity perturbation from the flow and the anisotropic stress, ${\Pi^{i}}_{j}$, is symmetric and traceless. These quantities are also decomposed according to (\ref{eqn:vecdecom}) and (\ref{eqn:tensdecom}). The scalar-vector-tensor (SVT) split of the perturbed Einstein equations then gives the following constraint and evolution equations.

\medskip
\noindent
Constraint:
\begin{subequations}
\begin{eqnarray}
 {\cal H} \dot{h} - 2 k^{2} \eta  &=& 8 \pi G a^{2} \delta \rho,\, \label{eqn:constraint1} \\
k \dot{\eta} &=&  4 \pi G a^{2} ( \rho + P ) v^{S}\,, \label{eqn:constraint2} \\
k \dot{H}^{V} &=& - 16 \pi G a^{2}( \rho + P ) v^{V}\,. \label{eqn:constraint3}
\end{eqnarray}
\end{subequations}
Evolution:
\begin{subequations}
\begin{eqnarray}
\ddot{h} + 2 {\cal H} \dot{h} - 2k^{2} \eta &=& - 24 \pi G a^{2} \delta P\,, \label{eqn:evolution1}  \\
\ddot{h} + 6 \ddot{\eta} + 2 {\cal H} ( \dot{h} + 6 \dot{\eta}) - 2 k^{2} \eta &=& - 16 \pi G a^{2} P \Pi^{S}\,, \label{eqn:evolution2} \\
\ddot{H}^{V}+2 {\cal H} \dot{H}^{V} &=&  8 \pi G a^{2} P \Pi^{V}\,, \label{eqn:evolution3} \\
 \ddot{H}^{T} + 2 {\cal H} \dot{H}^{T} + k^{2} H^{T} &=& 8 \pi G a^{2} P \Pi^{T}\,, \label{eqn:evolution4}
\end{eqnarray}
\end{subequations}
where $h=6H^{S}_{L}$ and $\eta=- \left(H^{S}_{L} + \frac{1}{3} H^{S}_{T} \right)$ are the metric variables defined in ref.~\cite{Ma:1995ey}. The equations of motion for the scalar and vector components of the displacement vector $\xi^{i}$ of the isotropic elastic fluid given by (\ref{eqn:solideom}) are then
\begin{subequations}
\begin{eqnarray}
\ddot{\xi}^{S}+ \left(1-3 \frac{dP}{d \rho}\right) {\cal H} \dot{\xi}^{S} +   k^{2} c_{\rm s}^{2} \left[ \xi^{S} + \frac{1}{2k} (h-h_{I}) \right] + 3k\left(c_{\rm s}^{2}-\frac{dP}{d \rho}\right)(\eta-\eta_{I}) &=& 0\,, \label{eqn:scalsdmeom} \\
\ddot{\xi}^{V} + \left(1-3 \frac{dP}{d \rho}\right) {\cal H} \dot{\xi}^{V} +  k^{2} c_{\rm v}^{2} \left[ \xi^{V} + \frac{1}{k}(H^{V}_{I}-H^{V}) \right]&=&0\,, \label{eqn:vecsdmeom}
\end{eqnarray}
\end{subequations}
where the subscript $I$ denotes the metric induced on the medium at the time of formation. Assuming an equation of state $P= w \rho$ then the scalar components of the perturbed energy-momentum tensor (\ref{eqn:T00elast}, \ref{eqn:Ti0elast}, \ref{eqn:Tijelast}) are, according to the decomposition (\ref{eqn:comT00}, \ref{eqn:comT0i}, \ref{eqn:comTi0}, \ref{eqn:comTij}), 
\begin{subequations}
\begin{eqnarray}
\delta \rho  &=& - \rho (1+w) \left[ k \xi^{S} + \frac{1}{2} (h-h_{I}) \right]\,, \label{eqn:rhoscal} \\
v^{S} &=& \dot{\xi}^{S}\,, \label{eqn:vscal} \\
\delta P &=& - \rho  (1+w) \frac{dP}{d \rho} \left[ k \xi^{S} + \frac{1}{2} (h-h_{I}) \right]\,, \label{eqn:pscal} \\
\Pi^{S} &=& \frac{3}{2}\left(c_{\rm s}^{2} - \frac{dP}{d \rho}\right)(1+w^{-1})  \left[ k \xi^{S} + \frac{1}{2} (h-h_{I}) +3(\eta-\eta_{I})\right] \\ \nonumber &=& \frac{3}{2}\left(c_{\rm s}^{2} - \frac{dP}{d \rho}\right)(1+w^{-1})  \left[ -\frac{\delta}{1+w} +3(\eta-\eta_{I})\right]\,. \label{eqn:piscal}
\end{eqnarray}
\end{subequations}
Note that $\delta P/\delta\rho=\textstyle{dP\over d\rho}$. Eqn.~(\ref{eqn:scalsdmeom}) can then be used in conjunction with (\ref{eqn:rhoscal}, \ref{eqn:vscal}, \ref{eqn:pscal}, \ref{eqn:piscal}) to give a closed set of scalar equations
\begin{subequations}
\begin{eqnarray}
\dot{\delta} &=& - (1+w) \left( k v^{S} + \frac{1}{2} \dot{h} \right)\,, \label{eqn:deltascaleqn} \\
\dot{v}^{S} &=& -{\cal H}\left(1-3 \frac{dP}{d \rho}\right) v^{S} + \frac{dP}{d \rho} \frac{1}{1+w} k \delta - \frac{2}{3} \frac{w}{1+w} k \Pi^{S}\,.  \label{eqn:vscaleqn}
\end{eqnarray}
\end{subequations}
The anisotropic stress is zero in the perfect fluid limit ($\mu=0$). Similarly, the vector sources are
\begin{subequations}
\begin{eqnarray}
v^{V} &=& \dot{\xi}^{V}\,, \label{eqn:vvec} \\
\Pi^{V} &=&  2 c_{\rm v}^{2} (1+ w^{-1})( k \xi^{V} + H^{V}_{I} - H^{V})\,, \label{eqn:pivec}  
\end{eqnarray}
\end{subequations}
and the anisotropic tensor source is
\begin{equation} \label{eqn:tenssource}
\Pi^{T} =  2 c_{\rm v}^{2} (1+ w^{-1}) ( H^{T}_{I} - H^{T})\,.
\end{equation}
The vector or tensor contributions due to the elastic medium are absent in the perfect fluid limit. Eqn.~(\ref{eqn:vecsdmeom}) can then be used in conjunction with (\ref{eqn:vvec}, \ref{eqn:pivec}) to give the vector equation of motion
\begin{equation} \label{eqn:veceom}
\dot{v}^{V}=-{\cal H} \left( 1-3 \frac{dP}{d \rho} \right) v^{V} - \frac{1}{2} \frac{w}{1+w} k \Pi^{V}\,.
\end{equation}

\subsection{Comparison to Generalized Fluid Systems} \label{sec:genfluid}

The equations of motion derived in the preceding sections are based on a very specific set of assumptions. Given that the understanding the evolution of dark energy perturbations is crucial to making precise predictions for the observed power spectra, a variety of phenomenological models have been discussed~\cite{Weller:2003hw,Bean:2003fb,Hu:1998kj,Hu:1998tj,Koivisto:2005mm}. In this section we attempt to make some contact between our work and these models.

Energy conservation of a generalized fluid energy-momentum tensor gives the scalar equations~\cite{Ma:1995ey,Hu:1998kj}:
\begin{subequations}
\begin{eqnarray}
\dot{ \left( \frac{\delta}{1+w} \right) } &=& - \left( k v^{S} + \frac{1}{2} \dot{h} \right)-3{\cal H} \frac{w}{1+w} \Gamma\,, \label{eqn:genfluid1} \\
\dot{v}^{S} &=& -{\cal H}\left(1-3 \frac{dP}{d \rho}\right) v^{S} + \frac{dP}{d \rho} \frac{1}{1+w} k \delta + \frac{w}{1+w} k \Gamma - \frac{2}{3} \frac{w}{1+w} k \Pi^{S}\,, \label{eqn:genfluid2}
\label{eqn:genfluid}
\end{eqnarray}
\end{subequations}
where $\Gamma$ is the entropy contribution which is given by 
\begin{equation}
 w \Gamma = \left( \frac{\delta P}{\delta \rho} - \frac{d P}{d \rho} \right) \delta\,.
\end{equation}
In general, $P= w \rho$ does not imply $\delta P = w \, \delta \rho$ due to temporal or spatial variations in $w$ which correspond to entropy perturbations. It transpires that both entropy and anisotropic stress can play a role in stabilizing perturbations which is crucial to the viability of any model.

In an elastic medium, we have already pointed out that $\delta\rho/\delta P=\textstyle{dP\over d\rho}$ which implies that $\Gamma=0$, that is, the medium is adiabatic. However, the non-zero rigidity leads to anisotropic stress and it is that which stabilizes perturbations as shown by~(\ref{eqn:soundconst1}).

A number of authors~\cite{Hu:1998kj,Hu:1998tj,Koivisto:2005mm} have suggested phenomenological approaches which involve the inclusion of anisotropic stress. Based on various arguments, they suggested making $\Pi^{S}$ dynamical and constructed a phenomenological equation of motion for its evolution
\begin{equation} \label{eqn:hustress}
\dot{\Pi}^{S} + {1\over T}\Pi^{S} = {4 c_{\rm vis}^{2}\over w} \left( k v^{S} + \frac{1}{2} \dot{h} + 3 \dot{\eta} \right)\,,
\end{equation}
where $T$ is some decay timescale (which they suggest should be $(3\mathcal{H})^{-1}$) and $c_{\rm vis}^2$ is some arbitrary coefficient, whose physical origin is attributed to viscosity within the fluid. If we differentiate (\ref{eqn:piscal}) and substitute in (\ref{eqn:deltascaleqn}), then we obtain
\begin{equation}
\dot{\Pi}^{S} =  \frac{2\mu}{P} \left( k v^{S} + \frac{1}{2} \dot{h} + 3 \dot{\eta} \right)\,,
\end{equation}
which has a similar form to (\ref{eqn:hustress}) if $T=\infty$ and $c_{\rm vis}^2=\mu/(2\rho)=c_{\rm v}^2(1+w)/2$. The construction of these phenomenological models was somewhat ad-hoc, nonetheless they are very close to those described here. However, we see that the physical mechanism which they describe is not viscosity, but rigidity. It might be possible to construct models with finite values of $T$ by modifying our approach to include some kind of dissipation.

We note that the system of equations which we have derived is very similar to that which might come from a truncated Boltzmann hierarchy for a black-body such as the neutrinos (or photons with no Thomson scattering terms). In particular, if we take $w=1/3$ and $\mu/\rho=4/15$, then the elastic model gives
\begin{subequations}
\begin{eqnarray}
\dot\delta&=&-{4\over 3}\left(kv^{S}+{1\over 2}\dot h\right)\,,\\
\dot v^{S}&=&k\left({1\over 4}\delta-{1\over 6}\Pi^{S}\right)\,,\\
\dot \Pi^{S}&=&{8\over 5}\left(kv^{S}+{1\over 2}\dot h+3\dot\eta\right)\,, 
\end{eqnarray}
\end{subequations}
which is exactly that given by a Boltzmann hierarchy with a third moment set to zero~\cite{Hu:1997mn}.

A second class of models seek to model dark energy fluids which are non-adiabatic and could come from scalar field models. For a minimally coupled scalar field with the standard kinetic term
\begin{equation}
\delta\rho=\dot\phi\dot{\delta\phi}+{dV\over d\phi}\delta\phi\,,\quad 
\delta P=\dot\phi\dot{\delta\phi}-{dV\over d\phi}\delta\phi\,,
\end{equation}
and hence $\delta P/\delta\rho$ clearly depends on the frame of the dark energy defined by $\delta\phi$ and $\dot{\delta\phi}$. One can choose to work in the rest frame of the dark energy defined by $(\rho+P)v^{i}={\hat k}^{i}k\delta\phi/\dot\phi=0$, where $c_{\rm s}^2=\delta P/\delta \rho=1$. More general models with non-minimal couplings or exotic kinetic terms could have $c_{\rm s}^2\ne 1$.

It was suggested in refs.~\cite{Weller:2003hw,Bean:2003fb} to use a system of equations described by~(\ref{eqn:genfluid1}, \ref{eqn:genfluid2}) with no anisotropic stress but with $\Gamma\ne 0$. In order to take into account the possibility of different frames, the sound speed was deemed to be defined in the rest frame of the dark energy and the equations of motion were modified to apply in an arbitrary frame, that is, one makes the transformation to
\begin{equation}
\delta^{{\rm rest}} = \delta + 3 {\cal H}(1+w)(v^{S}-B)/k \,.
\end{equation}
Here $B$ is the space-time component of the metric perturbation and is zero in both the synchronous and Newtonian gauges. This leads to equations of motion
\begin{subequations}
\begin{eqnarray}
\dot\delta&=&-(1+w)\left[kv^{S}\left(1+{9\mathcal{H}^2\over k^2}(c_{\rm s}^2-w)\right)+{1\over 2}\dot h\right]-3\mathcal{H}(c_{\rm s}^2-w)\delta\,,\\
\dot v^{S}&=&-\mathcal{H}(1-3 c_{\rm s}^2)v^{S}+{k c_{\rm s}^2\over 1+w}\delta\,,
\end{eqnarray}
\end{subequations}
which can be computed from (\ref{eqn:genfluid}) using 
\begin{equation}
w\Gamma_{\rm eff}=(c_{\rm s}^2-w)\left(\delta+3\mathcal{H}(1+w){v^{S}\over k}\right)\,.
\end{equation} 
When $c_{\rm s}^2=w$ these are clearly the same as for the elastic medium, but they are very different in the limit $c_{\rm s}^2\rightarrow 0$, as we shall see in the subsequent discussion.

\subsection{Anisotropic Generalizations} \label{sec:aniso}

In the previous sections we have described how one can construct a general perturbed energy-momentum tensor under a set of simple assumptions. The fact that the general isotropic tensor of rank four has a limited number of degrees of freedom has allowed us to construct the most general set of equations of motion for an isotropic medium defined by its density, pressure and rigidity. More generally we have shown that specification of {\it all} of the components of the pressure tensor,  $P^{\mu \nu}$, and elasticity tensor $E^{\mu \nu \rho \sigma}$ is sufficient to describe the perturbed energy momentum tensor using (\ref{eqn:emperturb}). In this section we will discuss aspects of a generalization which allows for the perturbations to be anisotropic and consider the case of cubic symmetry which is the simplest anisotropic possibility.

In general, the elasticity tensor has a total of 21 independent components~\cite{landau:1959}. One of these, the bulk modulus, is specified by the pressure and the other 20 are shear moduli, which can provide an anisotropic response to perturbations depending on the particular symmetry of the system. Fortunately the classification of these shear moduli has been studied in the context of classical elasticity theory and it is known that there 14 different types, known as the  Bravais lattices. In table~\ref{tab:noshearmod} we list the number of non-zero shear moduli for each symmetry group.

\renewcommand{\arraystretch}{1.6}
\begin{table}
\begin{center}
\begin{tabular}{|c|c|} \hline
 Symmetry & Number of Non-Zero Shear Moduli \\ \hline
    Triclinic & 20   \\
    Monoclinic & 12  \\
    Orthorhombic &  8  \\
    Tetragonal $({\bf C}_{\rm 4}, {\bf S}_{\rm 4}, {\bf C}_{\rm 4h})$ &  6  \\
    Tetragonal $({\bf C}_{\rm 4v}, {\bf D}_{\rm 2d}, {\bf D}_{\rm 4}, {\bf D}_{\rm 4h})$ & 5  \\
    Rhombohedral $({\bf C}_{\rm 3}, {\bf S}_{\rm 6})$ & 6  \\
    Rhombohedral $({\bf C}_{\rm 3v}, {\bf D}_{\rm 3}, {\bf D}_{\rm 3d})$ & 5  \\
    Hexagonal &  4 \\
    Cubic & 2  \\ 
Isotropic & 1  \\ \hline
\end{tabular}
\end{center}
\caption{\label{tab:noshearmod} Number of non-zero shear moduli for each of the Bravais lattice symmetry groups~\cite{landau:1959}. For the tetragonal and rhombohedral symmetries the number of shear moduli depend on the particular class of symmetry, which can be identified using the Sch\"{o}nflies notation. There are 7 shear moduli for tetragonal symmetry with a 4-fold rotation axis (${\bf C}_{\rm 4}$), for example, but this reduces to 6 with the additional of a mirror plane parallel to the axis of rotation (${\bf C}_{\rm 4v}$).}
\end{table}

An additional requirement that we will impose here is that the pressure tensor is isotropic, $P^{\mu\nu}=P\gamma^{\mu\nu}$, so that the unperturbed spacetime has the standard FRW metric, and the anisotropic response is only present at linearized order. This will probably restrict the range of possibilities allowed, but for sure there is at least one possibility, that of cubic symmetry, which is compatible with this.

The case of cubic response to perturbations was considered in ref.~\cite{Battye:2006mb}. There are now two shear moduli, $\mu_{\rm L}$ and $\mu_{\rm T}$, which specify the non-zero components of the  elasticity tensor
\begin{eqnarray}
E^{xxxx} &=& E^{yyyy} = E^{zzzz} = \beta+P+\frac{4}{3}\mu_{\rm L}\,, \\ \nonumber
E^{xxyy} &=& E^{yyzz} = E^{zzxx} = \beta-P-\frac{2}{3}\mu_{\rm L}\,, \\ \nonumber
E^{yzyz} &=& E^{xzxz} = E^{xyxy} = P+\mu_{\rm T}\,. 
\end{eqnarray}
The energy-momentum sources are modified from those in (\ref{eqn:T00elast}, \ref{eqn:Ti0elast}, \ref{eqn:Tijelast}) to
\begin{subequations}
\begin{eqnarray}
\delta T^{0}_{0} &=&  (\rho + P) \left( \partial_{i} \xi^{i} + \frac{1}{2} h \right)\,, \\
\delta T^{i}_{0} &=& -(\rho + P) \dot{\xi}^{i}\,, \\
\delta T^{i}_{j} &=& - \delta^{i}_{j} ( \beta - \frac{2}{3} \mu_{L} ) \left( \partial_{k} \xi^{k} + \frac{1}{2} h \right) - \mu_{L} (2 \partial_{(j} \xi^{i)} + h^{i}_{j}) - \Delta\mu \, {S^{i}}_{j}\,. 
\end{eqnarray}
\end{subequations}
where $\Delta\mu=\mu_{\rm T}-\mu_{\rm L}$ quantifies the degree of anisotropy. The tensor ${S^{i}}_{j}$ represents the cubic source term and is given by
\begin{equation}
{S^{i}}_{j} =  \left(  \begin{array}{ccc} 0 & 2 \partial_{(y} \xi^{x)} + h^{x}_{y} & 2 \partial_{(z} \xi^{x)} + h^{x}_{z} \\ 2 \partial_{(y} \xi^{x)} + h^{y}_{x} & 0  & 2 \partial_{(y} \xi^{z)} + h^{y}_{z} \\ 2 \partial_{(z} \xi^{x)} + h^{z}_{x}  & 2 \partial_{(z} \xi^{y)} + h^{z}_{y} & 0 \end{array} \right)\,.
\end{equation}
Furthermore, the evolution equations (\ref{eqn:solideom}) are modified to 
\begin{equation}
(\rho+P)(\ddot\xi^{i}+ {\cal H}\xi^{i})-3\beta {\cal H} \dot\xi^{i}-\beta (\partial^i\partial_j\xi^j+\partial^i h/2 ) -\mu_{\rm L} (\partial^j\partial_j\xi^i+\partial^i\partial_j\xi^j/3+\partial^j{h^{i}}_j-\partial^ih/3 )=\Delta\mu\,F^{i}\,,
\label{cubiceom}
\end{equation}
 where the cubic source term $F^{i}$ is given by 
\begin{equation}
F^{i} =  \left(  \begin{array}{ccc} ( \partial_{y} \partial^{y} + \partial_{z} \partial^{z}) \xi^{x} + \partial^{x} (\partial_{y} \xi^{y} + \partial_{z} \xi^{z}) + \partial^{y} {h^{x}}_{y} + \partial^{z} {h^{x}}_{z} \\  ( \partial_{x} \partial^{x} + \partial_{z} \partial^{z}) \xi^{y} + \partial^{y} (\partial_{x} \xi^{x} + \partial_{z} \xi^{z}) + \partial^{x} {h^{y}}_{x} + \partial^{z} {h^{y}}_{z} \\  ( \partial_{x} \partial^{x} + \partial_{y} \partial^{y}) \xi^{z} + \partial^{z} (\partial_{x} \xi^{x} + \partial_{y} \xi^{y}) + \partial^{x} {h^{z}}_{x} + \partial^{y} {h^{z}}_{y} \end{array} \right)\,.
\end{equation}
One of the consequences of the introduction of two shear moduli is that the sound speeds become dependent on the {\it direction} of the wave-vector~\cite{Battye:2005ik}. When one performs the same SVT decomposition as for the isotropic case, the different SVT sectors couple and scalar modes can excite vorticity and gravitational waves~\cite{Battye:2006mb}. It is is possible that this kind of anisotropic behaviour is responsible for anomalies seen the spectrum of the CMB on very large-scales; this is presently under investigation.

\section{Analytic Solutions} \label{sec:sdemodes}

In this section, we identify the regular perturbation modes which can arise in the early universe in the presence of an isotropic elastic medium. This task has been performed in the standard scalar case in ref.~\cite{Bucher:1999re}. Observations show that the primordial perturbation was most likely dominated by an adiabatic scalar mode generated from fluctuations in the metric. However, the most general primordial perturbation can also contain scalar isocurvature modes~\cite{Bucher:1999re} generated from variations in the abundance ratios of different particle species. In the vector sector, a regular mode can also be sourced from non-zero initial photon and neutrino vorticity~\cite{Lewis:2004kg} and primordial magnetic fields can also lead to regular modes~\cite{Lewis:2004ef}.

We consider a universe consisting of cold dark matter (c), baryons (b), neutrinos ($\nu$), photons ($\gamma$) and an elastic fluid component (e). We identify specific modes where the elastic medium has an equation of state with $w=0,-1/3,-2/3$. To identify the regular modes in the early universe we assume that the photons and baryons are tightly coupled due to the large Thomson scattering term $\sigma_{\rm T}$. One can then obtain exact equations for the evolution of their velocity. An expansion in opacity $\kappa_{\rm c}^{-1}=a n_{\rm e} \sigma_{\rm T}$ which is valid for max $( k \kappa_{\rm c}, {\cal H} \kappa_{\rm c} ) \ll 1$ gives
\begin{subequations}
\begin{eqnarray}
\dot{v}_{\rm \gamma b}^{S} ( 1 + R ) + R {\cal H} v_{\rm \gamma b}^{S} &=& \frac{k}{4} \delta_{\rm \gamma}\,, \\
\dot{v}^{V}_{\rm \gamma b} ( 1 + R ) + R {\cal H} v^{V}_{\rm \gamma b} &=& 0\,,
\end{eqnarray}
\end{subequations}  
where $R=3 \rho_{\rm b}/(4\rho_{\rm \gamma})$.  In the following we define $\omega_{\rm x}= \Omega_{\rm x} H_{0}^{2}$, $R_{\rm \nu}=\Omega_{\rm \nu}/\Omega_{\rm r}$ and $R_{\rm \gamma}=\Omega_{\rm \gamma}/\Omega_{\rm r}$. The primordial modes are then given by a series expansion of equations in Section~\ref{sec:svt_split} in terms of the conformal time $\tau$. 

\subsection{Scalar Modes} \label{sec:scalsdemodes}

In the scalar sector the most general primordial perturbation is specified by a mixture of adiabatic and isocurvature modes. Adiabatic coupling between various components in the universe requires that the relative perturbation between species, given by
\begin{equation}
A_{ij} = \frac{\delta_{i}}{1+w_{i}} - \frac{\delta_{j}}{1+w_{j}}\,,
\end{equation}
vanishes. If the various density perturbations compensate in such a way that the initial curvature perturbation ($\eta$) is zero on super-horizon scales then these are termed isocurvature initial conditions. The scalar adiabatic and isocurvature modes (for initial perturbations in fluids other than the elastic fluid) are listed in table~\ref{tab:adimodes}. 

The isocurvature modes which can exist due to an elastic fluid are listed in table~\ref{tab:isoscal}. These modes are interesting as they correspond to both  initial non-zero density fluctuations {\it and} non-zero anisotropic stress, resulting from fluctuations of the scalar component of the wordline displacement vector $\xi^S$. Physically, we expect these modes to arise due to a perturbed state of the elastic medium relative to the background at formation, that is, the medium is not formed in an equilibrium state.

Due to our definition of anisotropic stress, the stress term for an elastic fluid, defined in~(\ref{eqn:piscal}), will diverge when $w=0$. The equations of motion do not suffer any such divergence, as this arises only due to our use of the standard decomposition of the energy-momentum tensor in~(\ref{eqn:comTij}). As such, we list the non-zero initial term for the quantity  $w \, \Pi^{S}_{e}$ for the elastic anisotropic stress in table~\ref{tab:isoscal}.

\renewcommand{\arraystretch}{1.6}
\begin{sidewaystable} 
\begin{center}
\begin{tabular}{|c|c|c|c|c|c|} \hline
		& Adiabatic & CDM IC & Baryon IC & Neutrino Density IC & Neutrino Velocity IC \\ \hline
$h$		&   $\frac{1}{2} k^{2} \tau^{2}$ &  $\frac{\omega_{\rm c}}{\sqrt{\omega_{\rm r}}}\tau - \frac{3 \omega_{\rm c} (\omega_{\rm m}+\omega_{\rm e,w_{\rm e}=0})}{8 \omega_{\rm r}} \tau^{2}$ &   $\frac{\omega_{\rm b}}{\sqrt{\omega_{\rm r}}}\tau - \frac{3 \omega_{\rm b} (\omega_{\rm m}+\omega_{\rm e,w_{\rm e}=0})}{8 \omega_{\rm r}} \tau^{2}$   &  $\frac{R_{\rm \nu} \omega_{\rm b}}{40 R_{\rm \gamma} \sqrt{\omega_{\rm r}}} k^{2} \tau^{3}$      &  $\frac{3 R_{\rm \nu} \omega_{\rm b}}{8 R_{\rm \gamma} \sqrt{\omega_{\rm r}}} k \tau^{2}$                               \\ 

$\eta$		&  $1 -  \frac{(5 + 4 R_{\rm \nu})}{12(15 + 4 R_{\rm \nu})} k^{2} \tau^{2}$   &  $ - \frac{\rm \omega_{c}}{6\sqrt{\rm \omega_{r}}}\tau + \frac{\rm \omega_{c} \left[ (15+ 4 R_{\rm \nu}) \omega_{\rm m} + (5 + 4 R_{\rm \nu}) \omega_{\rm e,w_{\rm e}=0} \right] }{16 \omega_{\rm r} (15 + 4 R_{\rm \nu}) } \tau^{2}$   &    $ - \frac{\rm \omega_{b}}{6\sqrt{\rm \omega_{r}}}\tau + \frac{\rm \omega_{b} \left[ (15+ 4 R_{\rm \nu}) \omega_{\rm m} + (5 + 4 R_{\rm \nu}) \omega_{\rm e,w_{\rm e}=0} \right] }{16 \omega_{\rm r} (15 + 4 R_{\rm \nu}) } \tau^{2}$  &  $- \frac{R_{\rm \nu}}{6(15+4 R_{\rm \nu})} k^{2} \tau^{2} $                &   $- \frac{4 R_{\rm \nu}}{3(5+4 R_{\rm \nu})} k \tau $ \\ 

$\delta_{\rm c}$    &  $- \frac{1}{4} k^{2} \tau^{2}$ & $1 - \frac{ \omega_{\rm c}}{2\sqrt{\omega_{\rm r}}}\tau + \frac{3 \omega_{\rm c} (\omega_{\rm m}+\omega_{\rm e,w_{\rm e}=0}) }{16 \omega_{\rm r}} \tau^{2}$  &   $ - \frac{\omega_{\rm b}}{2\sqrt{\omega_{\rm r}}}\tau + \frac{3 \omega_{\rm b}  (\omega_{\rm m}+\omega_{\rm e,w_{\rm e}=0}) }{16 \omega_{\rm r}} \tau^{2}$   &   $-\frac{R_{\rm \nu} \omega_{\rm b}}{80 R_{\rm \gamma} \sqrt{\omega_{\rm r}}} k^{2} \tau^{3}$       &       $-\frac{3 R_{\rm \nu} \omega_{\rm b}}{16 R_{\rm \gamma} \sqrt{\omega_{\rm r}}} k \tau^{2}$                                 \\ 	

$\delta_{\rm b}$    &  $- \frac{1}{4} k^{2} \tau^{2}$ & $ - \frac{\omega_{\rm c}}{2\sqrt{\omega_{\rm r}}}\tau + \frac{3 \omega_{\rm c}  (\omega_{\rm m}+\omega_{\rm e,w_{\rm e}=0})}{16 \omega_{\rm r}} \tau^{2}$  &  $1 - \frac{\omega_{\rm b}}{2\sqrt{\omega_{\rm r}}}\tau + \frac{3 \omega_{\rm b}  (\omega_{\rm m}+\omega_{\rm e,w_{\rm e}=0})}{16 \omega_{\rm r}} \tau^{2}$    &   $ \frac{R_{\rm \nu}}{8 R_{\rm \gamma}} k^{2} \tau^{2}$     &      $\frac{R_{\rm \nu}}{R_{\rm \gamma}} k \tau - \frac{3(R_{\rm \gamma} + 2) R_{\rm \nu} \omega_{\rm b}}{16 R_{\rm \gamma}^{2} \sqrt{\rm \omega_{\rm r}}} k \tau^{2} $                               \\

$\delta_{\rm \gamma}$ &  $- \frac{1}{3} k^{2} \tau^{2}$ &  $- \frac{2 \omega_{\rm c}}{3\sqrt{\omega_{\rm r}}}\tau + \frac{ \omega_{\rm c}  (\omega_{\rm m}+\omega_{\rm e,w_{\rm e}=0})}{4 \omega_{\rm r}} \tau^{2}$   &    $- \frac{2 \omega_{\rm b}}{3\sqrt{\omega_{\rm r}}}\tau + \frac{ \omega_{\rm b}  (\omega_{\rm m}+\omega_{\rm e,w_{\rm e}=0})}{4 \omega_{\rm r}} \tau^{2}$     &  $- \frac{R_{\rm \nu}}{R_{\rm \gamma}} + \frac{R_{\rm \nu}}{6 R_{\rm \gamma}} k^{2} \tau^{2}$  &     $\frac{4 R_{\rm \nu}}{3 R_{\rm \gamma}} k \tau  - \frac{(R_{\rm \gamma} + 2) R_{\rm \nu} \omega_{\rm b}}{4 R_{\rm \gamma}^{2} \sqrt{\omega_{\rm r}}} k \tau^{2} $               \\ 

$\delta_{\rm \nu}$  &   $- \frac{1}{3} k^{2} \tau^{2}$  &  $- \frac{2 \omega_{\rm c}}{3\sqrt{\omega_{\rm r}}}\tau + \frac{ \omega_{\rm c}  (\omega_{\rm m}+\omega_{\rm e,w_{\rm e}=0})}{4 \omega_{\rm r}} \tau^{2}$   &   $- \frac{2 \omega_{\rm b}}{3\sqrt{\omega_{\rm r}}}\tau + \frac{ \omega_{\rm b}  (\omega_{\rm m}+\omega_{\rm e,w_{\rm e}=0})}{4 \omega_{\rm r}} \tau^{2}$     &      $1 - \frac{1}{6} k^{2} \tau^{2}$            &         $- \frac{4}{3} k \tau - \frac{R_{\rm \nu} \omega_{\rm b}}{4 R_{\rm \gamma} \sqrt{\omega_{\rm r}}} k \tau^{2} $            \\ 

$\delta_{\rm e, w_{\rm e}=0}$  & $- \frac{1}{4} k^{2} \tau^{2}$  & $- \frac{\omega_{\rm c}}{2\sqrt{\omega_{\rm r}}}\tau + \frac{3 \omega_{\rm c} \omega_{\rm m} }{16 \omega_{\rm r}} \tau^{2}$  & $- \frac{\omega_{\rm b}}{2\sqrt{\omega_{\rm r}}}\tau + \frac{3 \omega_{\rm b} \omega_{\rm m} }{16 \omega_{\rm r}} \tau^{2}$  &  $-\frac{R_{\rm \nu} \omega_{\rm b}}{80 R_{\rm \gamma} \sqrt{\omega_{\rm r}}} k^{2} \tau^{3}$   &       $-\frac{3 R_{\rm \nu} \omega_{\rm b}}{16 R_{\rm \gamma} \sqrt{\omega_{\rm r}}} k \tau^{2}$    \\ 

$\delta_{\rm e, w_{\rm e}=-1/3}$  &    $- \frac{1}{6} k^{2} \tau^{2}$ & $- \frac{ \omega_{\rm c}}{3\sqrt{\omega_{\rm r}}}\tau + \frac{ \omega_{\rm c} \omega_{\rm m}}{8 \omega_{\rm r}} \tau^{2}$   &   $- \frac{ \omega_{\rm b}}{3\sqrt{\omega_{\rm r}}}\tau + \frac{ \omega_{\rm b} \omega_{\rm m}}{8 \omega_{\rm r}} \tau^{2}$  &  $-\frac{R_{\rm \nu} \omega_{\rm b}}{120 R_{\rm \gamma} \sqrt{\omega_{\rm r}}} k^{2} \tau^{3}$   &      $-\frac{ R_{\rm \nu} \omega_{\rm b}}{8 R_{\rm \gamma} \sqrt{\omega_{\rm r}}} k \tau^{2}$    \\ 

$\delta_{\rm e, w_{\rm e}=-2/3}$   &   $- \frac{1}{12} k^{2} \tau^{2}$ &  $- \frac{ \omega_{\rm c}}{6\sqrt{\omega_{\rm r}}}\tau + \frac{ \omega_{\rm c} \omega_{\rm m}}{16 \omega_{\rm r}} \tau^{2}$   &    $- \frac{ \omega_{\rm b}}{6\sqrt{\omega_{\rm r}}}\tau + \frac{ \omega_{\rm b} \omega_{\rm m}}{16 \omega_{\rm r}} \tau^{2}$     &   $-\frac{R_{\rm \nu} \omega_{\rm b}}{240 R_{\rm \gamma} \sqrt{\omega_{\rm r}}} k^{2} \tau^{3}$   &      $-\frac{ R_{\rm \nu} \omega_{\rm b}}{16 R_{\rm \gamma} \sqrt{\omega_{\rm r}}} k \tau^{2}$    \\

$v^{S}_{\rm \gamma b}$ &  $- \frac{1}{36} k^{3} \tau^{3}$ &  $- \frac{\omega_{\rm c}}{12\sqrt{\omega_{\rm r}}} k \tau^{2}$   &   $- \frac{\omega_{\rm b}}{12\sqrt{\omega_{\rm r}}} k \tau^{2}$     &  $- \frac{R_{\rm \nu}}{4 R_{\rm \gamma}} k \tau + \frac{3}{16} \frac{R_{\rm \nu} \omega_{\rm b}}{R_{\rm \gamma} \sqrt{\omega_{\rm r}}}  k \tau^{2}$  &      $- \frac{R_{\rm \nu}}{R_{\rm \gamma}} + \frac{3 R_{\rm \nu} \omega_{\rm b}}{4 R_{\rm \gamma}^{2} \sqrt{\omega_{\rm r}}}  \tau$              \\ 

$v_{\rm \nu}$ &  $- \frac{(23 + 4 R_{\rm \nu})}{36(15 + 4 R_{\rm \nu})} k^{3} \tau^{3}$  &   $- \frac{\omega_{\rm c}}{12\sqrt{\omega_{\rm r}}} k \tau^{2}$   &   $- \frac{\omega_{\rm b}}{12\sqrt{\omega_{\rm r}}} k \tau^{2}$     &    $\frac{1}{4} k \tau - \frac{(27 + 4 R_{\rm \nu})}{72(15 + 4 R_{\rm \nu})} k^{3} \tau^{3} $              &     $1 - \frac{9 + 4 R_{\rm \nu}}{6(5 + 4 R_{\rm \nu})}  k^{2} \tau^{2}$               \\ 

$v^{S}_{\rm e, w_{\rm e}=0}$  & $- \frac{5 c_{\rm s}^{2}}{8(15 + 4 R_{\rm \nu})} k^{3} \tau^{3}$  &  $\frac{15 c_{\rm s}^{2}\omega_{\rm c} \omega_{\rm e}}{32(15+4 R_{\rm \nu})\omega_{\rm r}} k \tau^{3}$ &  $\frac{15 c_{\rm s}^{2}\omega_{\rm b} \omega_{\rm e}}{32(15+4 R_{\rm \nu})\omega_{\rm r}} k \tau^{3}$  &  $\frac{ c_{\rm s}^{2} R_{\rm \nu}}{8(15 + 4 R_{\rm \nu})} k^{3} \tau^{3}$  & $\frac{4 c_{\rm s}^{2} R_{\rm \nu}}{3(5 + 4 R_{\rm \nu})} k^{2} \tau^{2}$    \\ 

$v^{S}_{\rm e, w_{\rm e}=-1/3}$  &  $ \frac{(5 + 4R_{\rm \nu} - 30 c_{\rm s}^{2})}{60(15 + 4 R_{\rm \nu})}  k^{3} \tau^{3}$ &   $\frac{\omega_{\rm c}}{24\sqrt{\omega_{\rm r}}} k^{2} \tau^{2}$   &   $\frac{\omega_{\rm b}}{24\sqrt{\omega_{\rm r}}} k^{2} \tau^{2}$     &   $\frac{(3 c_{\rm s}^{2}+1) R_{\rm \nu}}{30(15 + 4 R_{\rm \nu})} k^{3} \tau^{3}$  &   $\frac{(3 c_{\rm s}^{2}+1) R_{\rm \nu}}{3(5 + 4 R_{\rm \nu})} k^{2} \tau^{2}$  \\ 

$v^{S}_{\rm e, w_{\rm e}=-2/3}$   &  $ \frac{(5 + 4R_{\rm \nu} - 15 c_{\rm s}^{2})}{36(15 + 4 R_{\rm \nu})}  k^{3} \tau^{3}$ &  $\frac{\omega_{\rm c}}{15\sqrt{\omega_{\rm r}}} k^{2} \tau^{2}$   &   $\frac{\omega_{\rm b}}{15\sqrt{\omega_{\rm r}}} k^{2} \tau^{2}$     &    $\frac{(3 c_{\rm s}^{2}+2) R_{\rm \nu}}{36(15 + 4 R_{\rm \nu})} k^{3} \tau^{3}$               &       $\frac{4(3 c_{\rm s}^{2}+2) R_{\rm \nu}}{15(5 + 4 R_{\rm \nu})} k^{2} \tau^{2}$              \\ 

$\Pi^{S}_{\rm \nu}$   &  $\frac{4}{(15+ 4 R_{\rm \nu})} k^{2} \tau^{2}$ & $-\frac{3}{15 + 4 R_{\rm \nu}} \omega_{\rm c} \omega_{\rm e} \tau^{2}, (w_{\rm e}=0) $     &    $-\frac{3}{15 + 4 R_{\rm \nu}} \omega_{\rm b} \omega_{\rm e} \tau^{2}, (w_{\rm e}=0) $     &    $\frac{3}{15 + 4 R_{\rm \nu}} k^{3} \tau^{3}$  &     $\frac{8}{(5+4 R_{\rm \nu})} k \tau$  \\

   &   & $-\frac{\omega_{\rm c}}{(15 + 2 R_{\rm \nu}) \sqrt{\omega_{\rm r}}} k^{2} \tau^{3}, (w_{\rm e}<0)$     &   $-\frac{\omega_{\rm b}}{(15 + 2 R_{\rm \nu}) \sqrt{\omega_{\rm r}}} k^{2} \tau^{3}, (w_{\rm e}<0)$       &                &               \\ \hline

\end{tabular}
\caption{\label{tab:adimodes} Regular adiabatic and isocurvature scalar modes. The adiabatic mode results from an initial curvature perturbation ($\eta$) and the isocurvature modes for CDM, baryons and neutrinos result from initial fluctuations in density, such that the curvature perturbation vanishes. There also exists a neutrino velocity mode, where the neutrino and photon-baryon fluid have a spatially varying relative velocity, balanced so that the net momentum is zero. We list perturbation growth in an elastic fluid for these modes.}
\end{center}
\end{sidewaystable}

\renewcommand{\arraystretch}{1.6}
\begin{table}
\begin{center}
\begin{tabular}{|c|c|c|c|} \hline 
	& $w_{\rm e}=0$ & $w_{\rm e}=-1/3$ & $w_{\rm e}=-2/3$  \\ \hline
$h$ &    $\frac{\omega_{\rm e}}{\sqrt{\omega_{\rm r}}} \tau$   &	$\frac{3}{4} \omega_{\rm e} \tau^{2}$ &  	$\frac{3}{5} \omega_{\rm e} \sqrt{\omega_{\rm r}} \tau^{3}$   \\

$\eta$	&  $ \left[  \frac{45 c_{\rm s}^{2} - 2(5 + 4 R_{\rm \nu})}{12(5+ 4 R_{\rm \nu})} \right] \frac{\omega_{\rm e}}{\sqrt{\omega_{\rm r}}} \tau$           & $\left[  \frac{30 c_{\rm s}^{2} - (5 + 4 R_{\rm \nu})}{8(15+ 4 R_{\rm \nu})} \right] \omega_{\rm e} \tau^{2}$ &  $\left[  \frac{75 c_{\rm s}^{2} - 2(5 + 4 R_{\rm \nu})}{40(15+ 2 R_{\rm \nu})} \right] \omega_{\rm e} \sqrt{\omega_{\rm r}} \tau^{3}$    \\ 

$\delta_{\rm c}$ &    $ -\frac{\omega_{\rm e}}{2\sqrt{\omega_{\rm r}}} \tau$              &  $- \frac{3}{8} \omega_{\rm e} \tau^{2}$  &  	$-\frac{3}{10} \omega_{\rm e} \sqrt{\omega_{\rm r}} \tau^{3}$   \\
 	
$\delta_{\rm b}$ &     $ -\frac{\omega_{\rm e}}{2\sqrt{\omega_{\rm r}}} \tau$          & $-\frac{3}{8} \omega_{\rm e} \tau^{2}$  &  	$-\frac{3}{10} \omega_{\rm e} \sqrt{\omega_{\rm r}} \tau^{3}$    \\

$\delta_{\rm \gamma}$ &    $- \frac{2\omega_{\rm e}}{ 3\sqrt{\omega_{\rm r}}} \tau$             & $- \frac{1}{2} \omega_{\rm e} \tau^{2}$ &  	$-\frac{2}{5} \omega_{\rm e} \sqrt{\omega_{\rm r}} \tau^{3}$    \\ 

$\delta_{\rm \nu}$  &  $-\frac{2\omega_{\rm e}}{3\sqrt{\omega_{\rm r}}} \tau$            &  $- \frac{1}{2} \omega_{\rm e} \tau^{2}$ & 	$-\frac{2}{5} \omega_{\rm e} \sqrt{\omega_{\rm r}} \tau^{3}$       \\ 

$\delta_{\rm e}$ & $1- \frac{\omega_{\rm e}}{2\sqrt{\omega_{\rm r}}} \tau$ &  $1 - \frac{1}{6} \left[ c_{\rm s}^{2} k^{2} + \frac{3}{2} \omega_{\rm e} \right] \tau^{2}$    &  $1 - \frac{1}{8} c_{\rm s}^{2} k^{2} \tau^{2}$    \\

$\xi^{S}_{\rm e}$ &  $ -k^{-1} \left[ 1 - \frac{1}{4} c_{\rm s}^{2} k^{2} \tau^{2} \right]$   &   $ -\frac{3}{2} k^{-1} \left[ 1 - \frac{1}{6} c_{\rm s}^{2} k^{2} \tau^{2} \right]$   &  $ -3 k^{-1} \left[ 1 - \frac{1}{8} c_{\rm s}^{2} k^{2} \tau^{2} \right]$      \\ 

$v^{S}_{\rm \gamma b}$ &   $-\frac{\omega_{\rm e}}{12\sqrt{\omega_{\rm r}}} k \tau^{2}$            & $-\frac{1}{24} \omega_{\rm e} k \tau^{3}$  &   $-\frac{1}{40} \omega_{\rm e}  \sqrt{\omega_{\rm r}} k \tau^{4}$    \\

$v^{S}_{\rm \nu}$ &  $-\frac{1}{12} \left[ \frac{18 c_{\rm s}^{2} + 5 + 4 R_{\rm \nu}}{5 + 4 R_{\rm \nu}} \right]  \frac{\omega_{\rm e}}{\sqrt{\omega_{\rm r}}} k \tau^{2}$              &  $-\frac{1}{24} \left[ \frac{ 24 c_{\rm s}^{2} + 23 + 4 R_{\rm \nu}}{15 + 4 R_{\rm \nu}} \right]  \omega_{\rm e} k \tau^{3}$        & $-\frac{1}{40} \left[ \frac{ 15 c_{\rm e}^{2} + 25+ 2 R_{\rm \nu}}{15 + 2 R_{\nu}} \right]  \omega_{\rm e}  \sqrt{\omega_{\rm r}} k \tau^{4}$   \\
 
$v^{S}_{\rm e}$ & $\frac{1}{2} c_{\rm s}^{2} k \tau  $   &  $\frac{1}{2} c_{\rm s}^{2} k \tau - \frac{1}{16} c_{\rm s}^{2} \frac{\omega_{\rm m}}{\sqrt{\omega_{\rm r}}} k \tau^{2}$       &  $\frac{3}{4} c_{s}^{2} k \tau - \frac{9}{80} c_{s}^{2} \frac{\omega_{\rm m}}{\sqrt{\omega_{\rm r}}} k \tau^{2}$        \\

$w \, \Pi^{S}_{\rm e}$   &  $-\frac{3}{2} c_{\rm s}^{2} + O(\tau)$   & $-\frac{1}{2} (3 c_{\rm s}^{2}+1)+ O(\tau)$    & $- (\frac{3}{2} c_{\rm s}^{2}+1)+ O(\tau)$   \\ 

$\Pi^{S}_{\rm \nu}$   &   $\frac{1}{2} \left[ \frac{3 c_{\rm s}^{2}}{5 + 4 R_{\rm \nu}} \right] \frac{\omega_{\rm e}}{\sqrt{\omega_{\rm r}}} \tau$              & $\frac{1}{6} \left[ \frac{3 c_{\rm s}^{2} + 1}{15 + 4 R_{\rm \nu}} \right] \omega_{\rm e} \tau^{2}$     &  $\frac{1}{12} \left[ \frac{3 c_{\rm s}^{2} + 2}{15 + 2 R_{\rm \nu}} \right] \omega_{\rm e} \sqrt{\omega_{\rm r}} \tau^{3}$    \\ \hline

\end{tabular}
\end{center}
\caption{\label{tab:isoscal} Isocurvature scalar modes due to an elastic fluid having an equation of state with $w=0$, $-1/3$ and $-2/3$. The isocurvature modes arise from spatially varying fluctuations in the elastic fluid resulting in  non-zero initial density fluctuations and anisotropic stress. We list the initial perturbation for the parameter combination $w \,\Pi^{S}_{\rm e}$, due to the definition of anisotropic stress in~(\ref{eqn:comTij}), which is regular even if $w=0$.}
\end{table}

\subsection{Vector Modes} \label{sec:vecsdemodes}

The regular vector mode has non-zero initial photon vorticity, having equal and opposite neutrino vorticity~\cite{Lewis:2004kg}. We have computed the initial conditions for this mode in the presence of a elastic fluid, which are shown in table \ref{tab:regvec}, neglecting the small contributions from the photon anisotropic stress. 

A regular mode also exists due to a non-zero initial vectorial component of the displacement vector $\xi^{V}$ in the elastic fluid, which also results in non-zero anisotropic stress (as in the scalar mode) and this mode is presented in table~\ref{tab:veciso}. In the elastic vector isocurvature mode the photon and neutrino vorticity are zero as they have no source term in the tight coupling limit. In both the regular vector mode and elastic isocurvature mode, the elastic fluid decouples from the equations of motion in the perfect fluid limit.

\renewcommand{\arraystretch}{1.6}
\begin{table}
\begin{center}
\begin{tabular}{|c|c|} \hline
 & Regular \\ \hline
$H^{V}$ &   $H^{V}_{1} \tau \left( 1 - \frac{15}{4} \frac{\omega_{\rm m} / \sqrt{\omega_{\rm r}}}{15 + 4 R_{\rm \nu}} \tau \right)$ \\ 
$v^{V}_{\rm \gamma b}$ & $- \frac{H^{V}_{1}}{4 k} \frac{4 R_{\rm \nu} + 5}{R_{\rm \gamma}} \left(1 - \frac{3}{4} \frac{ \omega_{\rm b} \sqrt{\omega_{\rm r}} }{\omega_{\rm \gamma}} \tau \right)$ \\
$v^{V}_{\rm \nu}$ & $\frac{H^{V}_{1}}{4 k} \frac{4 R_{\rm \nu} + 5}{R_{\rm \nu}}$ \\
$\Pi^{V}_{\rm \nu}$ & $\frac{2}{R_{\rm \nu}} H^{V}_{1} \tau$ \\
$\xi^{V}_{\rm e, w_{\rm e}=0}$ & $\frac{k c_{\rm v}^{2}}{9} H^{V}_{1} \tau^{3}$ \\
$\xi^{V}_{\rm e, w_{\rm e}=-1/3}$ & $\frac{k c_{\rm v}^{2}}{12} H^{V}_{1} \tau^{3}$ \\
$\xi^{V}_{\rm e, w_{\rm e}=-2/3}$ & $\frac{k c_{\rm v}^{2}}{15} H^{V}_{1} \tau^{3}$ \\ \hline
\end{tabular}
\end{center}
\caption{\label{tab:regvec} Regular vector mode with non-zero initial photon vorticity, having equal and opposite neutrino vorticity. $H^{V}_{1}$ is the first order term in the expansion for the metric perturbation $H^{V}$. We list perturbation growth in an elastic fluid for this mode.}
\end{table}

\renewcommand{\arraystretch}{1.6}
\begin{table}
\begin{center}
\begin{tabular}{|c|c|c|c|} \hline
	& $w_{\rm e}=0$ &  $w_{\rm e}=-1/3$ & $w_{\rm e}=-2/3$  \\ \hline

$H^{V}$ & $\left[ \frac{15 c_{\rm v}^{2}}{5 + 4 R_{\rm \nu}} \right] \frac{\omega_{\rm e}}{\sqrt{\omega_{\rm r}}}  \tau $      &  $\left[ \frac{10 c_{\rm v}^{2}}{15 + 4 R_{\rm \nu}} \right] \omega_{\rm e}  \tau^{2} $    &   $\frac{1}{2} \left[ \frac{5 c_{\rm v}^{2}}{15 + 2 R_{\rm \nu}} \right] \omega_{\rm e} \sqrt{\omega_{\rm r}}  \tau^{3} $     \\ 

$\xi^{V}_{\rm e}$ &  $ k^{-1} \left[ 1 - \frac{1}{4} c_{\rm v}^{2} k^{2} \tau^{2} \right] $   &   $ k^{-1} \left[ 1 - \frac{1}{6} c_{\rm v}^{2} k^{2} \tau^{2} \right] $   &  $ k^{-1} \left[ 1 - \frac{1}{8} c_{\rm v}^{2} k^{2} \tau^{2} \right] $  \\ 

$v^{V}_{\rm \nu}$ &  $\frac{1}{2} \left[ \frac{3 c_{\rm v}^{2}}{5 + 4 R_{\rm \nu}} \right] \frac{\omega_{\rm e}}{\sqrt{\omega_{\rm r}}} k \tau^{2} $  &  $\frac{1}{3} \left[ \frac{2 c_{\rm v}^{2}}{15 + 4 R_{\rm \nu}} \right] \omega_{\rm e} k \tau^{3} $          &   $\frac{1}{8} \left[ \frac{ c_{\rm v}^{2}}{15 + 2 R_{\rm \nu}} \right] \omega_{\rm e} \sqrt{\omega_{\rm r}} k \tau^{4} $    \\ 

$w \, \Pi^{V}_{\rm e}$  &  $2 c_{\rm v}^2 \left[ 1 - \left( \frac{15 c_{\rm v}^{2}}{5 + 4 R_{\rm \nu}} \right) \frac{\omega_{\rm e}}{\sqrt{\omega_{\rm r}}} \tau  \right]$  & $\frac{4}{3} c_{\rm v}^2 \left[ 1- \left( \frac{c_{\rm v}^{2} (5 k^{2} + 4 R_{\rm \nu} k^{2} + 60 \omega_{\rm e} )}{6(5 + 4 R_{\rm \nu})} \right)  \tau^{2}  \right]$  &  $\frac{2}{3} c_{\rm v}^2 \left[ 1- \frac{1}{8} c_{\rm v}^{2} k^{2} \tau^{2} \right]$ \\ 

$\Pi^{V}_{\rm \nu}$ &  $-\left[ \frac{24 c_{\rm v}^{2}}{5 + 4 R_{\rm \nu}} \right] \frac{\omega_{\rm e}}{\sqrt{\omega_{\rm r}}} \tau $     &   $-\left[ \frac{16 c_{\rm v}^{2}}{15 + 4 R_{\rm \nu}} \right] \omega_{\rm e} \tau^{2} $     &  $-\left[ \frac{4 c_{\rm v}^{2}}{15 + 2 R_{\rm \nu}} \right] \omega_{\rm e} \sqrt{\omega_{\rm r}} \tau^{3} $    \\ \hline

\end{tabular}
\end{center}
\caption{\label{tab:veciso} Isocurvature vector modes due to an elastic fluid having an equation of state with $w=0$, $-1/3$ and $-2/3$. This mode is the analogue of the scalar isocurvature mode, with a non-zero initial vectorial component of the displacement vector resulting in non-zero anisotropic stress.}
\end{table}

The vector isocurvature solution for a $w=-1/3$ elastic component can be found analytically in the radiation era due to the absence of photon and neutrino vorticity. This solution is given by
\begin{equation} \label{stringsolution}
\xi^{V}= \xi^{V}_{I} \left[ \left(1-\frac{4 \omega_{\rm e}}{k^{2}+ 4 \omega_{\rm e}} \right) \frac{\sin \left[ c_{\rm v} (k^{2}+ 4 \omega_{\rm e})^{\frac{1}{2}} \tau \right]}{c_{\rm v} (k^{2}+ 4 \omega_{\rm e})^{\frac{1}{2}} \tau} + \frac{4 \omega_{\rm e}}{k^{2}+ 4 \omega_{\rm e}} \right]\,.
\end{equation}
It can be shown that this solution gives that presented in table~\ref{tab:veciso} if $\xi^{V}_{I}=k^{-1}$ when expanded as a power series in $\tau$. 

\subsection{Tensor Modes}

There are no tensor modes other than the standard adiabatic mode. The tensor solution can be found analytically for a $w=-1/3$ elastic component in the radiation era assuming that the anisotropic stress of photons and neutrinos is zero. Using (\ref{eqn:evolution4}) and (\ref{eqn:tenssource}) then 
\begin{equation} \label{eqn:tenssolution}
H^{T}=H_{I}^{T}\left[\frac{\sin \left[ (k^{2}+4 c_{\rm v}^{2}\omega_{\rm e} )^{\frac{1}{2}} \tau \right] k^{2}}{(k^{2}+4 c_{\rm v}^{2}\omega_{\rm e} )^{\frac{3}{2}} \tau} + \frac{4 c_{\rm v}^{2}\omega_{\rm e} }{k^{2}+4 c_{\rm v}^{2}\omega_{\rm e} } \right]\,.
\end{equation}
This reverts to the standard solution $H^{T}=H_{I}^{T} j_{0} (k \tau)$ when $\omega_{\rm e}=0$ or $c_{\rm v}^{2}=0$.

\section{Cosmological Signatures} \label{sec:cosmo}

The observable CMB and matter power spectra were computed by modifying the CAMB software~\cite{Lewis:1999bs} to include an elastic fluid. We consider the elastic fluid in two scenarios - in one instance it acts as the dark energy component in an otherwise standard cosmology. The shear modulus $\mu$ stabilizes perturbations when $w<0$. We also consider in models with $\Omega_{\Lambda}\ne0$ and the elastic fluid as a pressureless component with $w=0$, but with a non-zero shear modulus. The shear modulus introduces a clustering scale, the Jeans length $\lambda_{\rm J}\sim c_{\rm s} \tau_{0}$, where $\tau_{0}$ is the conformal time today, as $c_{\rm s}^{2} = 4 \mu/(3 \rho)$ when $w=0$. This shares some similarities with a hot (or warm) dark matter component as power will be suppressed on small scales, although the shear modulus also allows for vector perturbations in the medium. 

\subsection{Scalar Sector} \label{sec:scalobser}

The scalar $C_{\ell}$'s are given by
\begin{equation}
C_{\ell} = 4 \pi  \int d (\log k) \mathcal{P}_{s}(k) |\Delta_{\ell}(k,\tau_{0})|^{2}\,,
\end{equation}
where $\Delta_{\ell}(k,\tau_{0})$ is the associated multipole moment for the photon distribution and $\mathcal{P}_{s}(k)$ is the initial power spectrum, parameterized by $\mathcal{P}_{s}(k)=A_{\rm s} k^{n_{\rm s}-1}$, where $A_{\rm s}$ is the initial scalar amplitude and $n_{\rm s}$ is the scalar spectral index.
Note that $\mathcal{P}(k)=k^{3} P(k)/(2\pi^{2})$, where $P(k)$ is the matter power spectrum.

\subsubsection{Adiabatic Mode}

The adiabatic mode is likely to have dominated the primordial fluctuation as a good fit to both CMB and galaxy clustering data can be obtained by this mechanism of structure formation and, assuming a flat universe, specifying a total of 6 cosmological parameters - $\Omega_{\rm b}$, $\Omega_{\rm c}$, $A_{\rm s}$, $n_{\rm s}$, $h$ and the optical depth to reionization $\kappa_{\rm R}$. A dark energy component is required in order to reconcile the observed accelerated expansion~\cite{Perlmutter:1996ds,Riess:1998cb,Perlmutter:1998np,Riess:2001gk,Astier:2005qq}.

The introduction of a dark energy component effects the CMB anisotropies both by its influence on the background expansion rate and its gravitational perturbations. Any component which has the same equation of state $w$ will give an identical overall expansion effect. However, an elastic fluid can potentially be distinguished by the different evolution of perturbations compared with other dark energy models. If $w<0$ then small-scale anisotropies are unaffected by the dark energy component as the perturbations at these scales entered the horizon when the fractional dark energy density was negligible. The primary contribution of the dark energy then arises at large angular scales through the Integrated Sachs-Wolfe (ISW) effect. The temperature-temperature (TT) contribution of the ISW effect to the scalar multipoles is given by
\begin{equation}
C_{\ell}^{TT}= 4 \pi  \int d (\log k) \mathcal{P}_{s}(k) \left[ \int_{\tau_{\rm dec}}^{\tau_{\rm 0}} d \tau e^{- \kappa} (\dot{\phi} + \dot{\psi}) j_{\ell} (k(\tau_{\rm 0} -\tau)) \right]^{2}\,,
\end{equation}
where $\phi$ and $\psi$ are the Newtonian potentials~\cite{Ma:1995ey} and $\tau_{\rm dec}$ is the conformal time at decoupling. The presence of anisotropic stress alters the evolution of the Newtonian potentials. 

\begin{figure}[] 
\begin{center}
\raisebox{2.7cm}{\resizebox{0.30\textwidth}{!}{\includegraphics{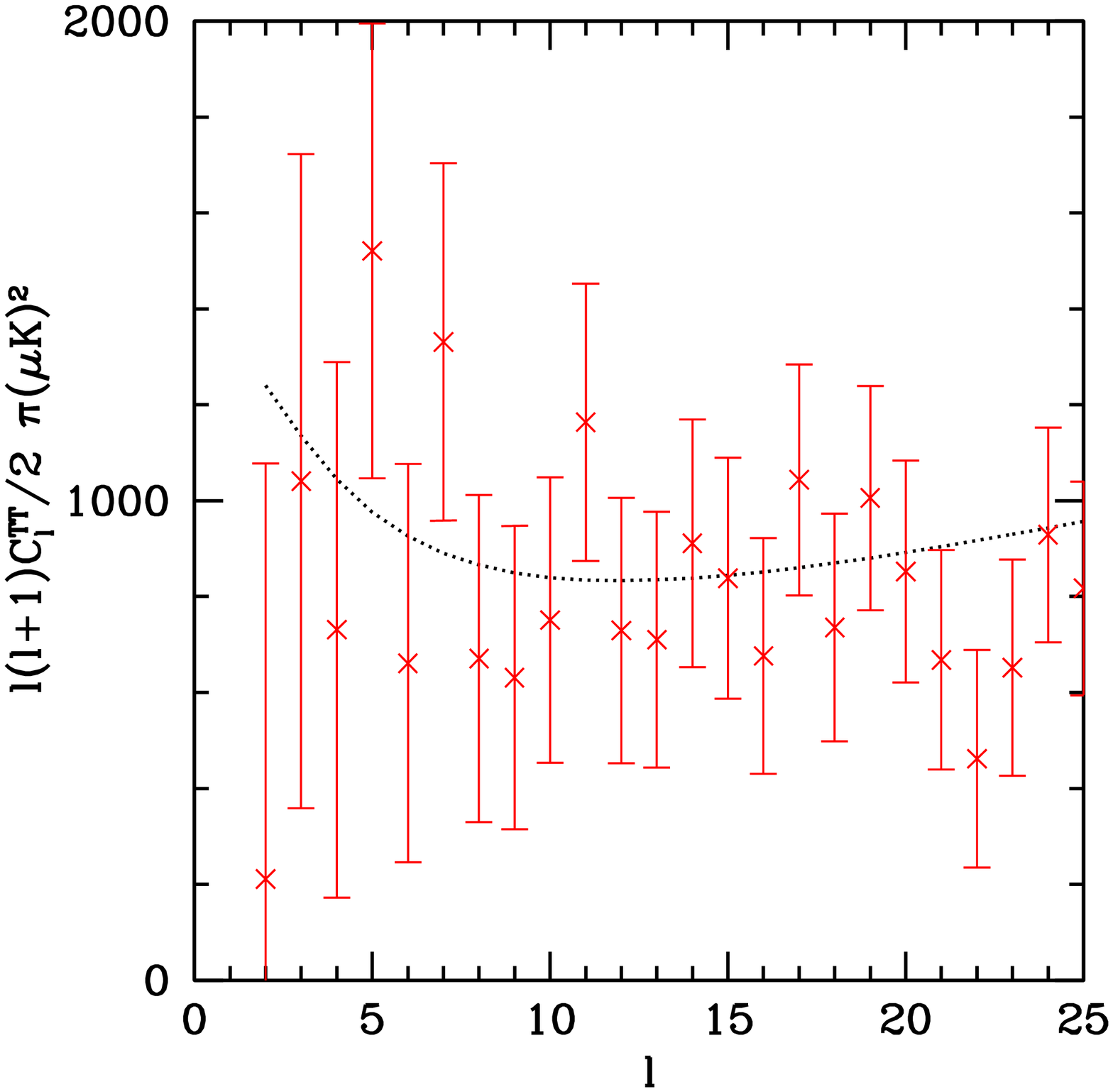}}}
\mbox{\resizebox{0.60\textwidth}{!}{\includegraphics{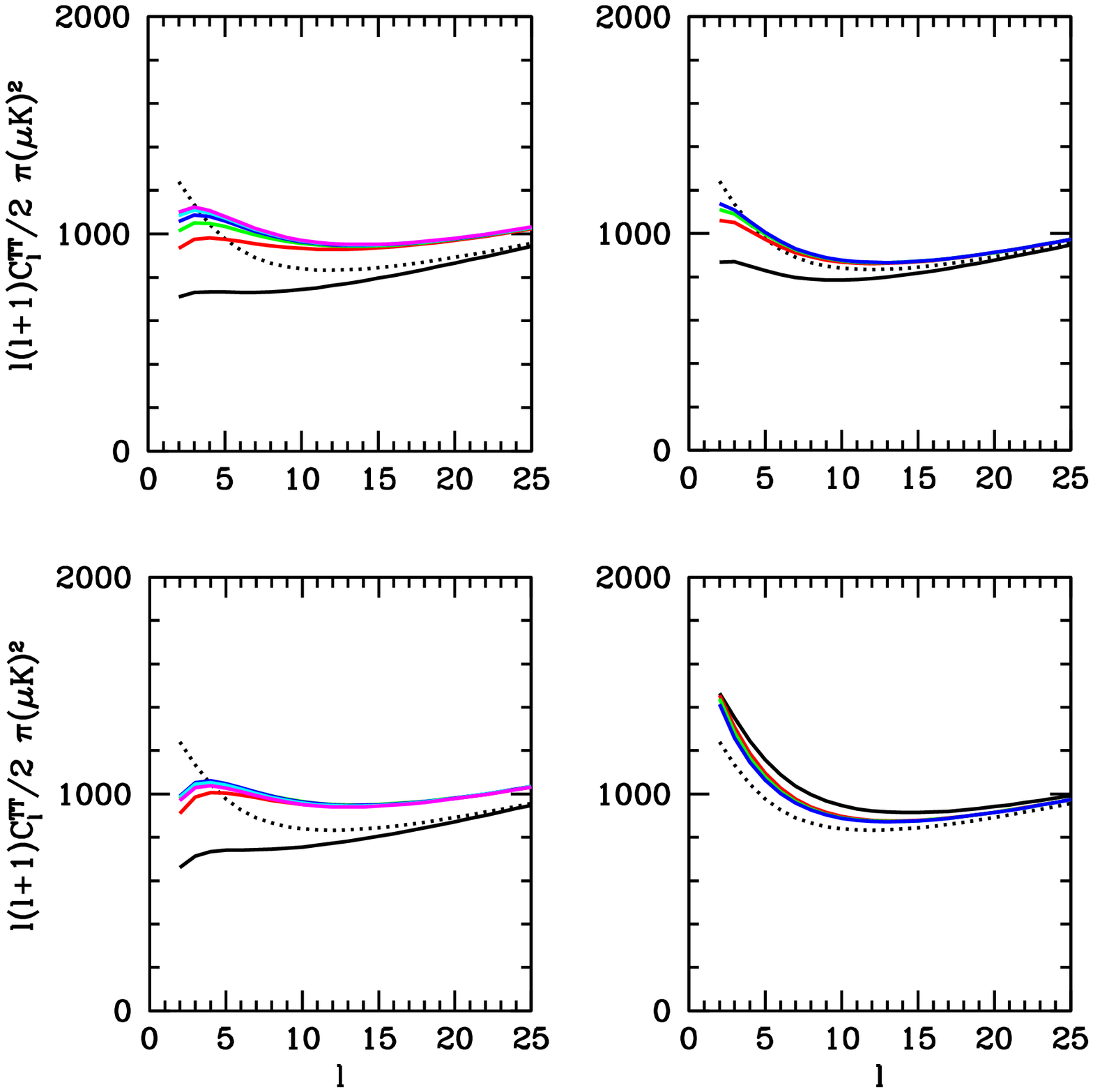}}}
\caption{\label{fig:lambdacdmcmbTT} (Left) A comparison of CMB TT data with the best-fitting WMAP $\Lambda$CDM model at low multipoles. The parameters used are $\Omega_{\rm c} h^2 = 0.104$, $\Omega_{\rm b} h^{2}= 0.0223$, $h=0.734$, $n_{\rm s}=0.951$, $A_{\rm s} = 2.02 \times 10^{-9}$ and $\kappa_{\rm R} = 0.088$~\cite{Spergel:2006hy}. 
(Right set of 2 $\times$ 2) A comparison between scalar field and elastic fluid dark energy models at low multipole values. The upper panels show scalar field models and the lower elastic models. The left and right panels show values of $w=-1/3$ and $w=-2/3$ respectively. In each case the angular diameter distance to the first acoustic peak has been rescaled to coincide with the best-fit $\Lambda$CDM model to isolate the ISW effect. The appropriate change to the Hubble parameter is $h= \{0.50,\,0.60\}$  for $w=\{-1/3,\,-2/3\}$. In each diagram a dotted line shows the $\Lambda$CDM model. 
The solid curves (bottom to top apart from the $w=-2/3$ elastic fluid model) are $c_{s}^{2}=0$, 0.2, 0.4, 0.6, 0.8, 1, where $c_{\rm s}^{2} \le 2/3$ in the $w=-2/3$ case due to the stability condition~(\ref{eqn:soundconst1}).} 
\end{center}
\end{figure}

In Fig.~\ref{fig:lambdacdmcmbTT} we plot the CMB anisotropies for multipoles $\ell < 25$ for the best fitting $\Lambda$CDM model~\cite{Spergel:2006hy}, along with the best estimates of the TT WMAP three year data~\cite{Hinshaw:2006ia,Page:2006hz}. We also plot the anisotropies for a number of elastic fluid models, along with the corresponding scalar field model, for various values of $w$ and $c_{\rm s}^{2}$. In each case we exploit the degeneracy between $w$ and $h$ to rescale the first acoustic peak in order to isolate the ISW effect. It is noticeable in both dark energy models that there is a reduction in power at large angular scales for $w=-1/3$ - this reduction is slightly more pronounced for the elastic model due to the anisotropic stress. For the scalar field models, the ISW effect is minimised for $c_{\rm s}^{2}=0$ and thereby increases monotonically with $c_{\rm s}^{2}$. The elastic model is consistent with this behaviour for $w=-1/3$, but for $w=-2/3$ $c_{\rm s}^{2}=0$ actually maximizes the ISW effect.

\begin{figure}[] 
\begin{center}
\raisebox{2.7cm}{\resizebox{0.30\textwidth}{!}{\includegraphics{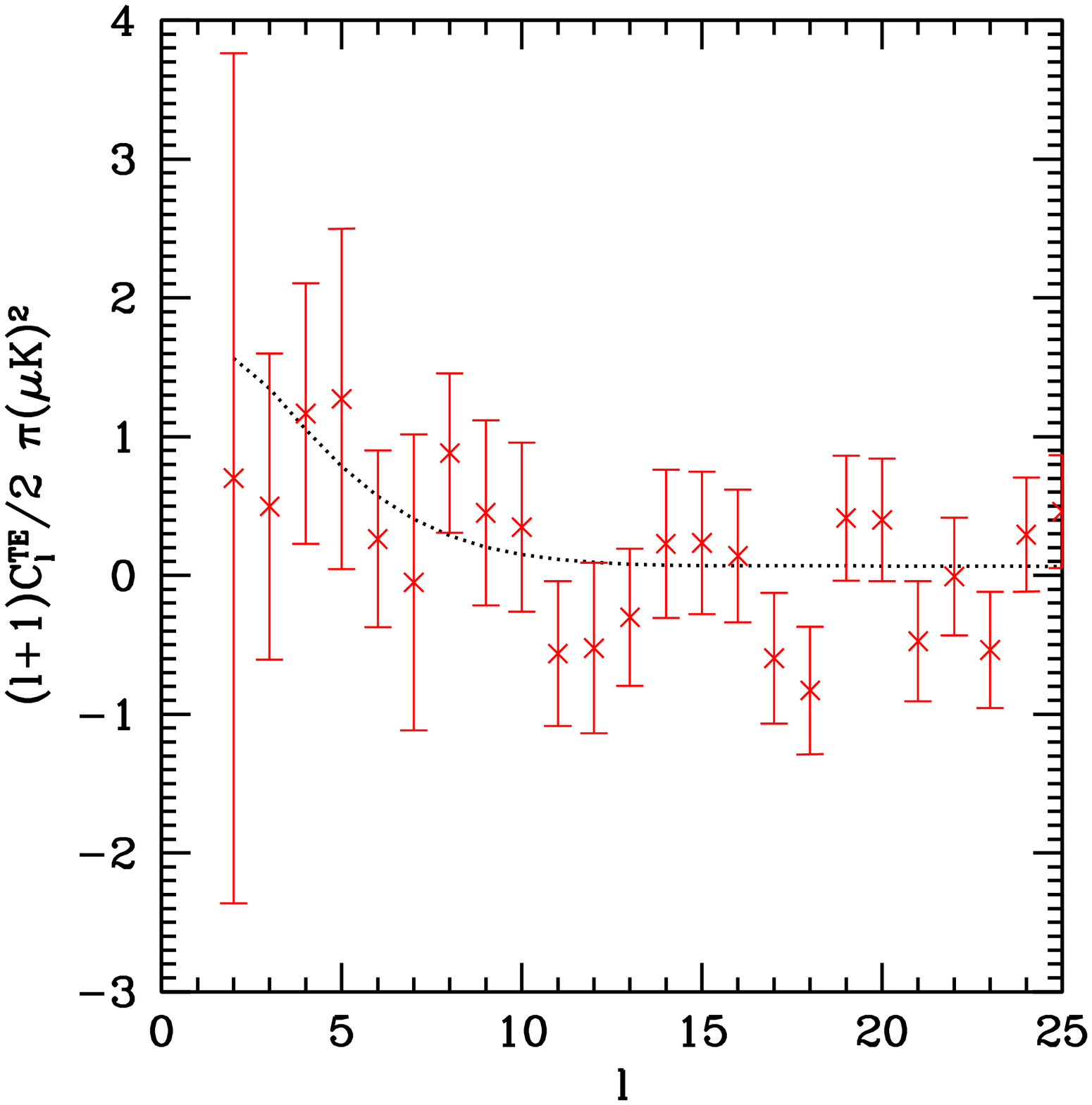}}}
\mbox{\resizebox{0.60\textwidth}{!}{\includegraphics{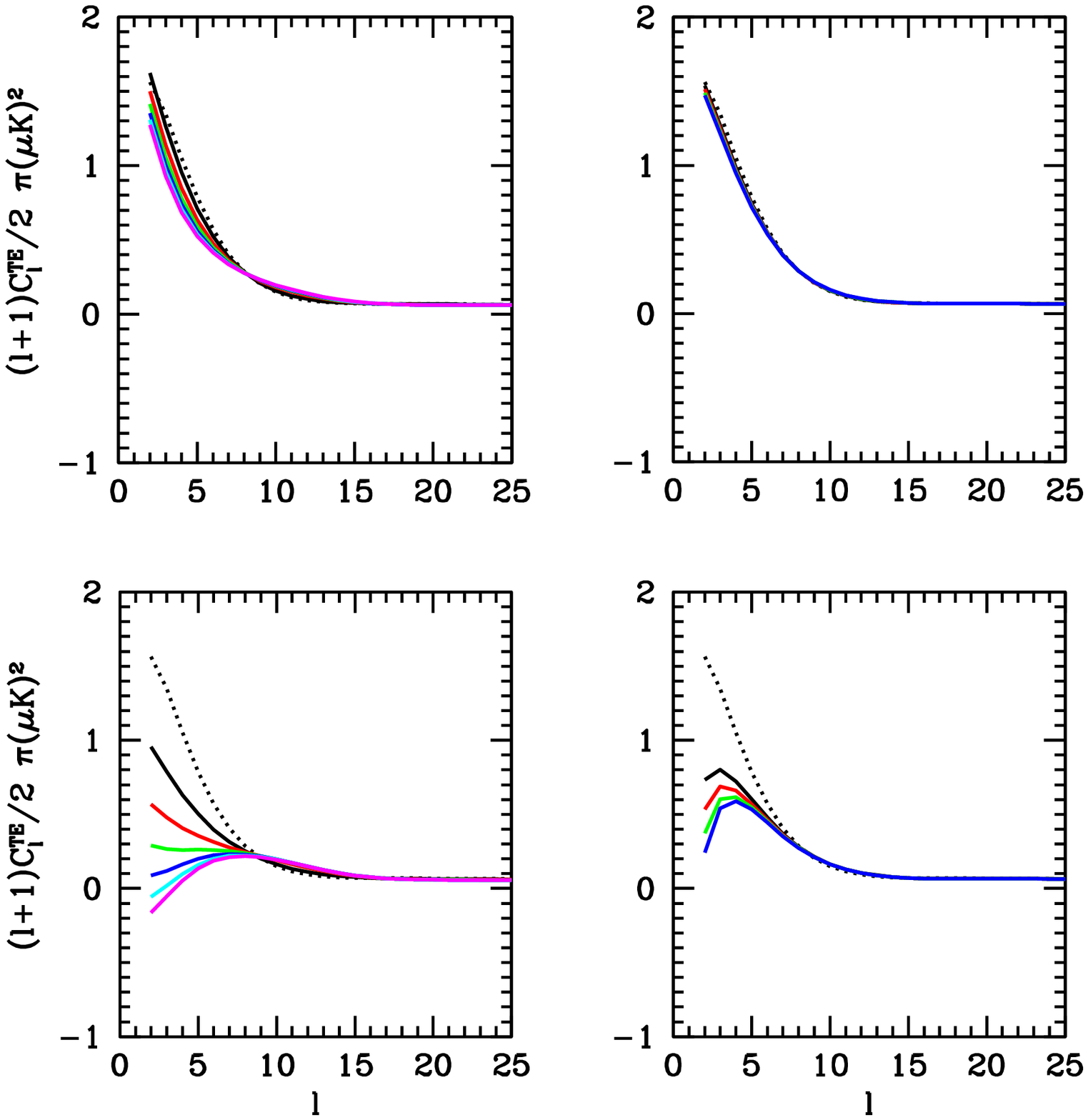}}}
\caption{\label{fig:lambdacdmcmbTE} (Left) A comparison of CMB TE data with the best-fitting WMAP $\Lambda$CDM model at low multipoles. (Right set of 2 $\times$ 2) A comparison between scalar field and elastic fluid dark energy models at low multipole values. Parameter values and labelling is the same as in Fig~\ref{fig:lambdacdmcmbTT}, apart from $c_{s}^{2}$ where (top to bottom) $c_{s}^{2}=0$, 0.2, 0.4, 0.6, 0.8, 1 and $c_{\rm s}^{2} \le 2/3$ in the $w=-2/3$ case due to the stability condition~(\ref{eqn:soundconst1}).}  
\end{center}
\end{figure}

In Fig.~\ref{fig:lambdacdmcmbTE} we plot the temperature-polarization (TE) power spectrum for the best fitting $\Lambda$CDM model. The signal on large angular scales can be attributed to early reionization. The contribution to the large scale TE power for the polarization generated at reionization is given by 
\begin{eqnarray}
C_{\ell}^{TE} &=& 4 \pi  \int d (\log k) \mathcal{P}_{s}(k) \left[ \int_{\tau_{\rm re}}^{\tau_{\rm 0}} d \tau e^{- \kappa} (\dot{\phi} + \dot{\psi}) j_{\ell} (k(\tau_{\rm 0} -\tau)) \right] \\ \nonumber &\times& \left[ \int_{\tau_{\rm re}}^{\tau_{\rm 0}} - d \tau \frac{3}{4k^{2}} (g(k^{2} \Pi + \ddot{\Pi})+2\dot{g}\dot{\Pi}+\ddot{g}\Pi) j_{\ell} (k(\tau_{\rm 0} -\tau)) \right]\,,
\end{eqnarray}
where $g=\dot{\kappa}\,e^{- \kappa}$ is the visibility function, $\tau_{\rm re}$ is the conformal time at reionization and the polarization source is given in terms of the photon anisotropic stress and the zeroth and second order polarization moments by $\Pi=\Pi_{\gamma}^{S}/3 + \Delta_{P0} + \Delta_{P2}$. The photon anisotropic stress is the dominant term coming from the free streaming of the monopole at recombination. We also plot the TE power spectrum for a number of elastic fluid models, along with the corresponding scalar field models, for various values of $w$ and $c_{\rm s}^{2}$. We find that precision measurement of the TE cross correlation at low $\ell$ can provide a potentially powerful discriminator of the dark energy model, if non-zero anisotropic stress is present. The variation of $c_{\rm s}^{2}$ for a scalar field model generating no anisotropic stress has little effect on the TE spectrum. The elastic fluid, however, shows significant differences for scales of $\ell<10$ due to anisotropic stress modifying the decay of gravitational potentials during reionization.

\begin{figure}[] 
\centering
\mbox{\resizebox{0.7\textwidth}{!}{\includegraphics{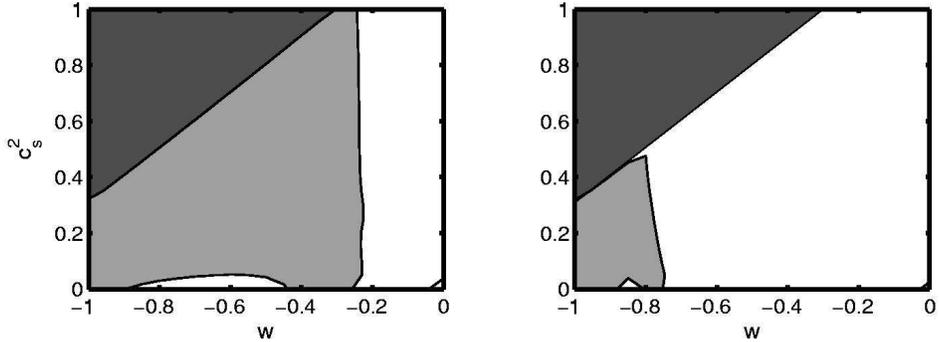}}}
\caption{\label{fig:discrimfig} The discrimination `landscape' between scalar field and elastic dark energy based on CMB TT (left panel) and TE (right panel) data. The dark grey shaded area shows the region of parameter space excluded by the bound $c_{\rm v}^{2}<1$ for the elastic component, and the light grey area the region where the two models cannot be distinguished. It becomes difficult to distinguish the models in the region of cosmological interest between $ -1 \le w \le -1/3$.}
\end{figure}

The possibility of distinguishing two competing models whose differences are only significant on large angular scales is limited by the effect of cosmic variance. The probability of distinguishing model A from B, assuming that A is correct, is then given by~\cite{Battye:1999eq} 
\begin{equation}
\left \langle \ln \left( \frac{P(\{a_{\ell m} \} | A)}{P(\{a_{\ell m} \} | B)}\right) \right \rangle_{A} = -\frac{1}{2} \sum_{\ell} (2\ell + 1) \left[ 1- \frac{C_{\ell}^{(A)}}{C_{\ell}^{(B)}} + \ln \left( \frac{C_{\ell}^{(A)}}{C_{\ell}^{(B)}} \right) \right]. 
\end{equation}
In Fig.~\ref{fig:discrimfig} we plot this quantity for $A=$ a scalar field model and $B=$ an elastic fluid model in the $(w,c_{s}^{2})$ parameter space, and indicate the region where the models can be distinguished at 1$-\sigma$  . We find that this probability is too small in almost all of the region of interest to discriminate between models, based on CMB TT anisotropies. In the right panel of Fig.~\ref{fig:discrimfig} we plot the analogous quantity based on TE anisotropies, assuming $\kappa_{\rm R}\ne0$, which shows that accurate measurement at $\ell<10$ vastly increases the area of parameter space that can be differentiated. If $\kappa_{\rm R}=0$ one would be unable to distinguish between scalar field models from elastic models using TE.

\begin{figure}[]
\begin{center}
\mbox{\resizebox{0.49\textwidth}{!}{\includegraphics{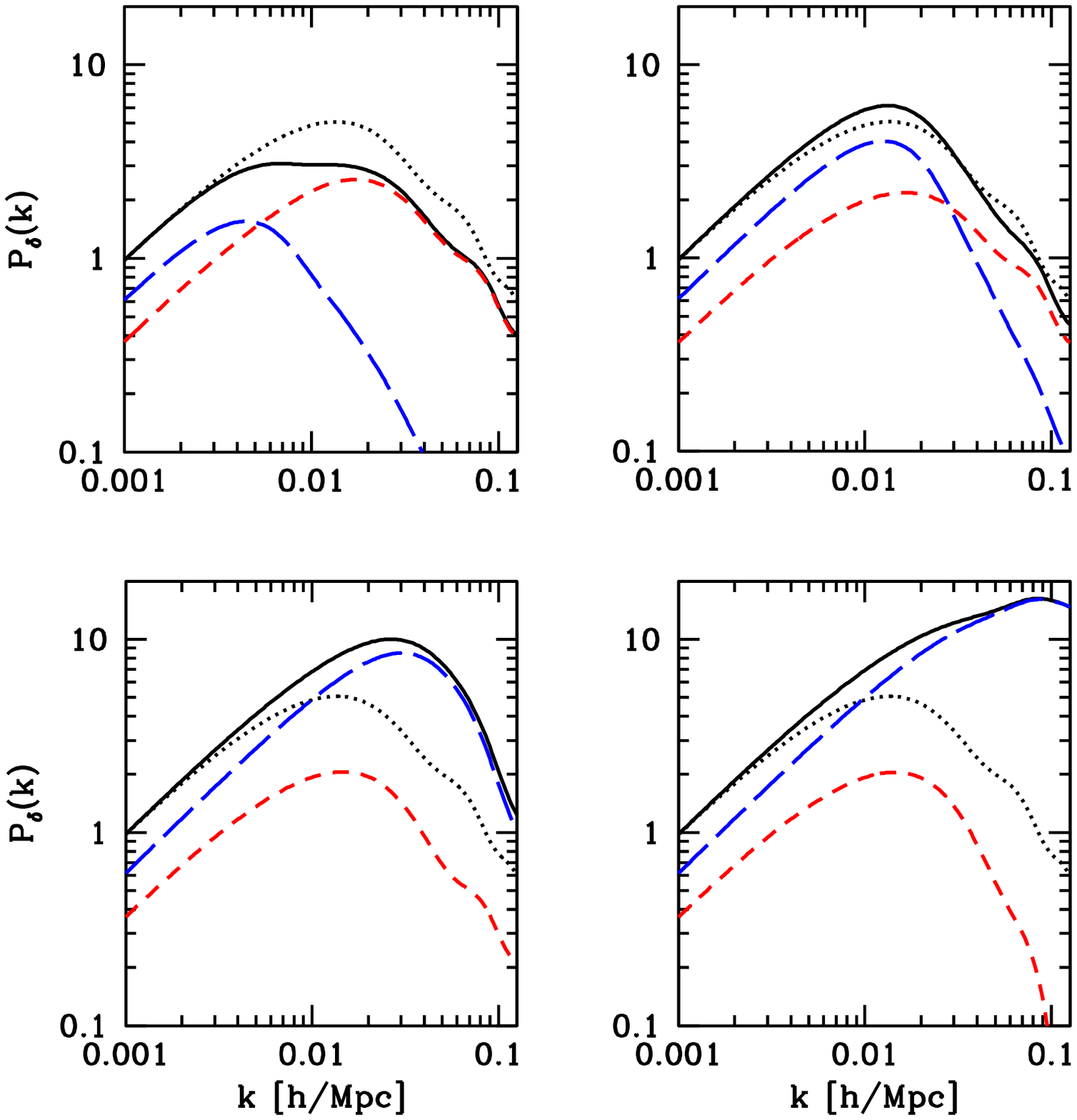}}}
\mbox{\resizebox{0.49\textwidth}{!}{\includegraphics{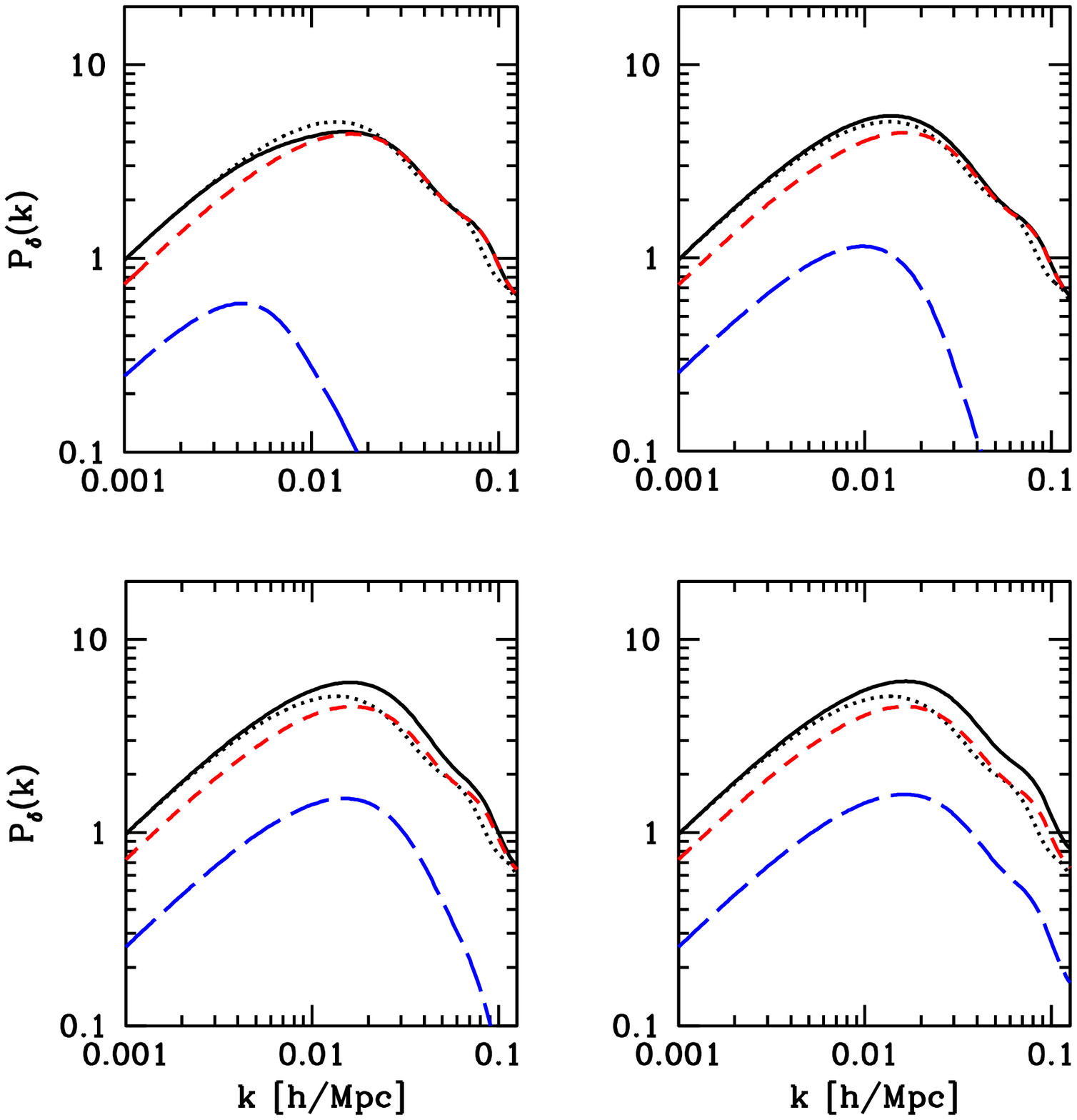}}}
\caption{\label{fig:adisolidmpk} The clustering effect of elastic (left set of $2 \times 2$) and scalar field (right set of $2 \times 2$) models as $c_{\rm s}^{2} \rightarrow 0$. The solid curve shows the total power spectrum for $w=-2/3$, while the long and short-dashed curves show the decomposition of the total power into components due to the clustering of the dark energy and CDM/baryonic species respectively. The panels show a dark energy sound speed of $c_{\rm s}^{2}=10^{-2}$ (top-left), $c_{\rm s}^{2}=10^{-3}$ (top-right),  $c_{\rm s}^{2}=10^{-4}$ (bottom-left) and  $c_{\rm s}^{2}=10^{-5}$ (bottom-right). The dotted curve shows the $\Lambda$CDM model. The total power spectrum has been arbitrarily normalized at $k=10^{-3} h {\,\rm Mpc^{-1}}$.} 
\end{center}
\end{figure}

The large scale structure of the universe depends on the growth rate of perturbations in the various species. An important aspect of dark energy is that it has a large Jeans length, $\lambda_{\rm J}\sim c_{\rm s} \tau_{0}$, at the present epoch so that it does not cluster on small scales and contribute to measurements of $\Omega_{\rm m}$ in galaxy clusters. We have assumed that the matter power spectrum is modified to 
\begin{equation} \label{eqn:matterpower}
P_{\delta}(k) = b^{2} |\delta_{\rm T} |^{2}\,,
\end{equation}
where $\delta_{\rm T}=\sum_{i}\Omega_{i}\delta_{i}$. This makes this untested assumption that the biasing of the dark energy component is proportional to its density. In Fig.~\ref{fig:adisolidmpk} we plot the matter power spectrum for $w=-2/3$ elastic and scalar field dark energy models, decomposing the power into components due to the various species. 

There are noticeable differences in the two cases. The presence of a clustering scale in the dark energy ($c_{\rm s}^2 <<1$) affects the elastic fluid power spectrum in several ways. At small scales, power increases as the sound speed becomes non-relativistic. If the assumption~(\ref{eqn:matterpower}) is correct then an approximate limit of $c_{\rm s}^{2} \gtrsim 10^{-3}$ would be required in order to be compatible with large scale structure (LSS) data. It is also noticeable that power in the CDM/baryonic components is reduced as the sound speed of the elastic component becomes non-relativistic. In the scalar field dark energy model, however, a sound speed as low as $c_{\rm s}^{2}=10^{-5}$ does not change the total matter power spectrum significantly. We have reduced $c_{\rm s}^{2}$ further and find that $c_{\rm s}^{2}=0$ would be compatible with LSS data. 

\begin{figure}[] 
\centering
\mbox{\resizebox{0.35\textwidth}{!}{\includegraphics{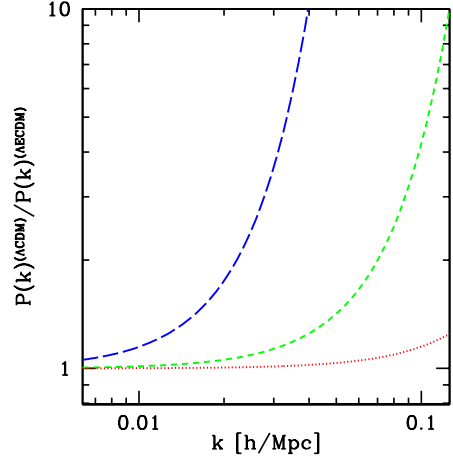}}}
\caption{\label{fig:ecdm} The ratio of the $\Lambda$CDM matter power spectrum and models where the CDM has a non-zero shear modulus, $\Lambda$ECDM, with sound speeds of $c_{s}^{2}=10^{-6}$ (dotted line), $c_{s}^{2}=10^{-5}$ (short-dash) and $c_{s}^{2}=10^{-4}$ (long-dash).} 
\end{figure}

Finally, we also consider whether the introduction of a non-zero shear modulus in a pressureless $w=0$ component is compatible with CMB and LSS data. In Fig.~\ref{fig:ecdm} we plot the ratio of the $\Lambda$CDM matter power spectrum  against models where there is elastic rigidity in the CDM, which we denote $\Lambda$ECDM. The introduction of a non-zero shear modulus introduces a clustering scale in the CDM and power is suppressed on small scales. When $c_{\rm s}^{2}=10^{-5}$, for example, power is suppressed by an order of magnitude compared to the $\Lambda$CDM model at $k=0.1 h {\,\rm Mpc^{-1}}$. For the values of $c_{\rm s}^{2}<10^{-4}$, the CMB anisotropies are affected at less than the $1 \%$ level by variations in $c_{\rm s}^{2}.$

\subsubsection{Isocurvature Modes}

\begin{figure}[] 
\centering
\mbox{\resizebox{0.325\textwidth}{!}{\includegraphics{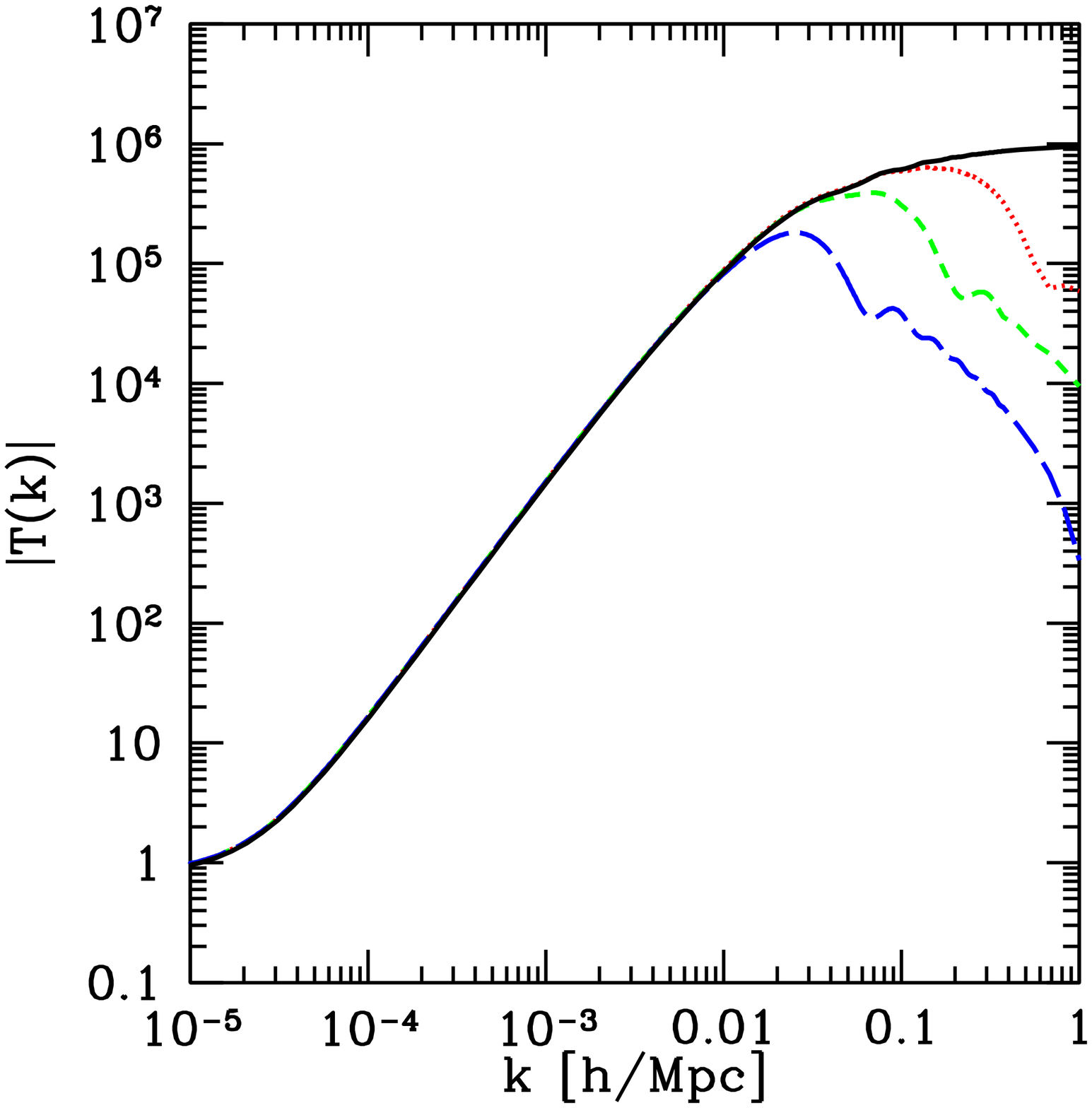}}}
\mbox{\resizebox{0.325\textwidth}{!}{\includegraphics{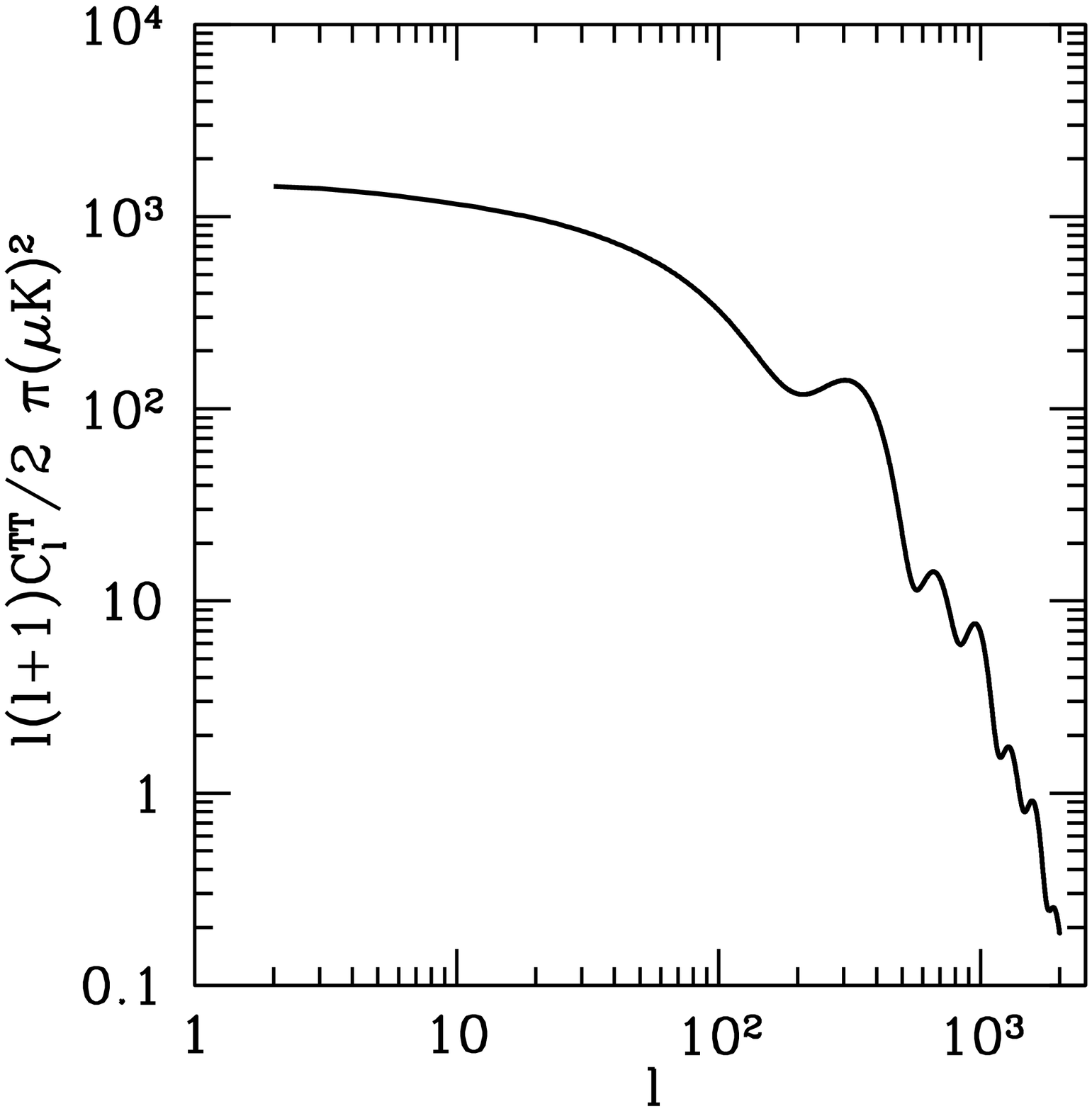}}}
\mbox{\resizebox{0.325\textwidth}{!}{\includegraphics{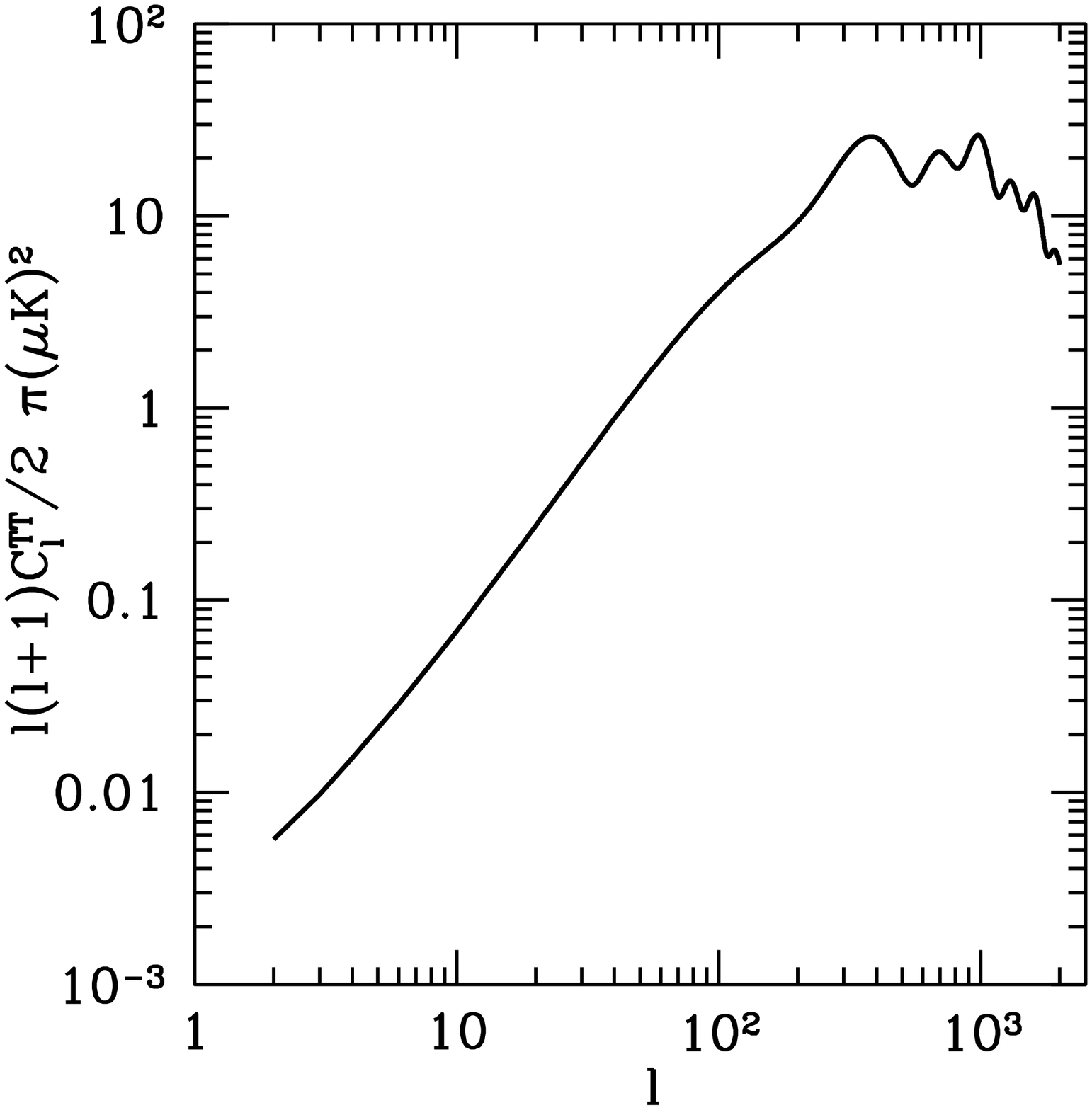}}}
\caption{\label{fig:transiso} Transfer function of the total density fluctuation $\delta_{\rm T}$ (left), along with the scale-invariant (middle) and white noise (right) CMB power spectrum for the $w=0$ elastic model with isocurvature initial fluctuations. The sound speeds are $c_{\rm s}^{2}=0$ (solid line), $c_{\rm s}^{2}=10^{-6}$ (dotted), $c_{\rm s}^{2}=10^{-5}$ (short-dash) and $c_{\rm s}^{2}=10^{-4}$ (long-dash). Transfer functions have been arbitrarily normalised to unity at $k=1\times 10^{-5}  h \, {\rm Mpc}^{-1}$.} 
\end{figure}

\begin{figure}[] 
\centering
\mbox{\resizebox{0.7\textwidth}{!}{\includegraphics{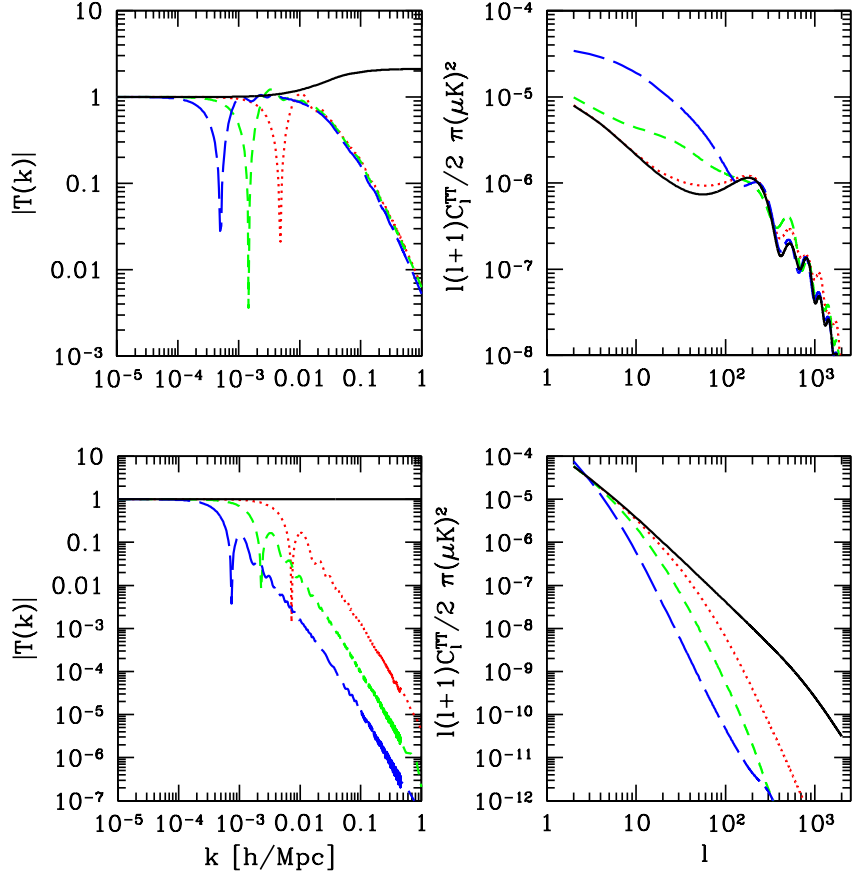}}}
\caption{\label{scalsdm} Transfer function (left) and white noise CMB power spectrum (right) for $w=-1/3$ (top) and $-2/3$ (bottom) with isocurvature initial fluctuations. The sound speeds are $c_{\rm s}^{2}=0$ (solid line), $c_{\rm s}^{2}=0.005$ (dotted), $c_{\rm s}^{2}=0.05$ (short-dash) and $c_{\rm s}^{2}=0.5$ (long-dash).} 
\end{figure}

Although the initial conditions appear to be dominated by the adiabatic mode, a sub-dominant isocurvature contribution cannot be ruled out. The scale-invariant CDM isocurvature mode suffers from the problem that the CMB anisotropy has greatly suppressed power on small scales relative to large scales. The reason for this is that the Sachs-Wolfe effect and the initial temperature fluctuation both add to give a much greater contribution to large-scale power. Recent work has placed an approximate 10$\%$ upper bound on the CDM isocurvature fraction, and similar constraints also apply for the other isocurvature modes~\cite{Bean:2006qz}.

The power spectrum of the scalar isocurvature mode is defined by
\begin{equation}
\langle |\delta|^{2} \rangle = \int d(\log k) \, \mathcal{P}_{\delta}(k).
\end{equation}
The scalar isocurvature modes for an elastic fluid are listed in table~\ref{tab:isoscal}. We list modes for a pressureless $w=0$ component with a non-zero shear modulus, which can be compared to the standard CDM isocurvature mode, in table~\ref{tab:adimodes}. The effect of the shear modulus, and hence the sound speed $c_{\rm s}^{2}$, enters at first order in $\tau$ for several of the variables, such as the metric perturbation $\eta$. We also list isocurvature modes where the elastic fluid has an equation of state $w=-1/3$ and $-2/3$.

In Fig.~\ref{fig:transiso} we plot the transfer function $T(k)$ of the total density fluctuation $\delta_{\rm T}$, along with the CMB anisotropies for a scale-invariant ($\mathcal{P}_{\delta} \propto k^{0}$, corresponding to $P_{\delta}(k)\propto k^{-3}$) and white noise power spectrum ($\mathcal{P}_{\delta} \propto k^{3}$, corresponding to $P_{\delta}(k)\propto k^{0}$) for $w=0$ and a primordial power amplitude ratio of $\mathcal{P}_{\delta}/\mathcal{P}_{\eta} \sim 1$, where $\mathcal{P}_{\eta}$ is the power in the curvature perturbation. As in the adiabatic case, density fluctuations are suppressed on small scales when $c_{s}^{2} \ne 0$. Since the curvature fluctuation is zero initially, there is the characteristic phase shift of the CMB peak positions typical of isocurvature modes compared to the adiabatic one, with the first angular peak at $\ell \sim 350$. For the values of $c_{\rm s}^{2}<10^{-4}$, we again find that the CMB anisotropies are affected at less than the $1 \%$ level. When $w<0$ perturbation growth is suppressed and $T(k)$ is much flatter, with a fall off in power when $c_{\rm s}^{2} \ne 0$. This is because as these modes cross through the horizon the dark energy density is small, and so the growth in the curvature fluctuation is supressed. These are shown in Fig.~\ref{scalsdm}, where we also plot the CMB anisotropies for a white noise power spectrum when $w=-1/3$ and $-2/3$.
 
\subsection{Vector Sector} \label{sec:vecobser}

Primordial vector modes constitute vortical perturbations in the early universe. The regular vector mode has been discussed in ref.~\cite{Lewis:2004kg}, along with a mode sourced by the anisotropic stress of a primordial magnetic field in ref~\cite{Lewis:2004ef}. We have also shown in section~\ref{sec:vecsdemodes} that a regular solution exists with a non-zero wordline displacement $\xi^{V}$, which corresponds to non-zero initial anisotropic stress in the elastic fluid. 

\subsubsection{Regular Mode}

\begin{figure}[] 
\centering
\mbox{\resizebox{0.4\textwidth}{!}{\includegraphics{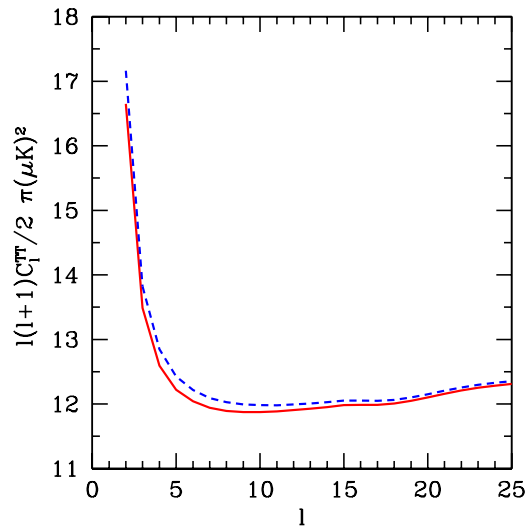}}}
\caption{CMB TT power spectrum at low multipoles for the regular vector mode with a scale invariant spectrum and primordial vector to scalar power ratio of $\sim 10^{-3}$. The dashed curve shows both a $\Lambda$CDM model {\em and} an elastic fluid with $w=-1/3$ and $c_{\rm v}^{2}=0$ (corresponding to $c_{\rm s}^{2}=-1/3$). The solid curve shows an elastic fluid with $w=-1/3$ and $c_{\rm v}^{2}=c_{\rm s}^{2}=1$.}
\label{regvec}
\end{figure}

In ref.~\cite{Lewis:2004kg} vector perturbations are expressed in the zero vorticity frame, which coincides with the synchronous gauge used here. The Einstein equation was written
\begin{equation} \label{eqn:vector2}
k (\dot{\sigma} + 2 \mathcal{H} \sigma) = - 8 \pi G a^{2} \Pi^{V},
\end{equation}
where $\sigma$ is the harmonic coefficient of the shear tensor. The relationship $\dot{H}^{V}=-k \sigma$ gives the Einstein evolution equation~(\ref{eqn:evolution3}), and substituting $H^{V}_1=-k \sigma_{0}$ into the series solution in section~\ref{sec:vecsdemodes} returns the same result as in ref.~\cite{Lewis:2004kg}. The primordial power spectrum is defined by
\begin{equation} 
\langle |\sigma|^{2} \rangle = \int d(\log k) \, \mathcal{P}_{\sigma}(k).
\end{equation}
In Fig.~\ref{regvec} we plot the CMB TT anisotropies for multipoles $\ell < 25$ for a scale invariant spectrum ($\mathcal{P}_{\sigma} \propto k^{0} $) with a primordial power amplitude ratio of $\mathcal{P}_{\sigma}/\mathcal{P}_{\eta} \sim 10^{-3}$. 

When the vector sound speed is zero the elastic dark energy does not contribute to the anisotropic stress of the Einstein equation~(\ref{eqn:vector2}). We find that the large scale CMB power is then the same for both the $\Lambda$CDM model and elastic dark energy models with $c_{\rm v}^{2}=0$. However, $w<0$ requires a {\em non-zero} vector sound speed otherwise the scalar perturbations will be unstable to collapse, as shown by~(\ref{eqn:soundconst1}). In the $w=-1/3$ case, for example, $c_{\rm v}^{2} \ge 1/4$ is required for $c_{\rm s}^{2} \ge 0$. However, the effect of the anisotropic stress on the CMB anisotropies at large angular scales is relativity small, as shown in Fig.~\ref{regvec}.

\subsubsection{Isocurvature Modes}

\begin{figure}[] 
\centering
\mbox{\resizebox{0.4\textwidth}{!}{\includegraphics{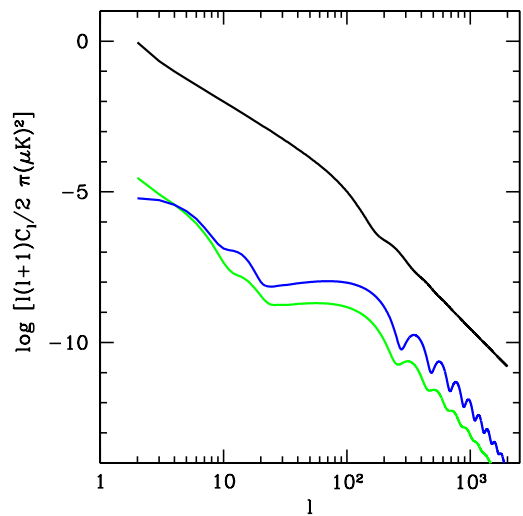}}}
\caption{CMB anisotropies for the vector isocurvature mode with a white noise $(\mathcal{P}_{\Pi} \propto k^{3})$ spectrum, $w=-1/3$ and $c_{\rm s}^{2}=1/2$ ($c_{\rm v}^{2}=5/8$). The top curve is the temperature spectrum and below this the B- and E-mode polarization.}
\label{fig:isovecn4}
\end{figure}

Isocurvature modes can exist in an elastic fluid due to non-zero anisotropic stress. We quantity these modes by the spectrum
\begin{equation} 
\langle |\Pi|^{2} \rangle = \int d(\log k) \, \mathcal{P}_{\Pi}(k).
\end{equation}
In Fig.~\ref{fig:isovecn4} we plot an example of this mode for $w=-1/3$, with a white noise power spectrum and a primordial power amplitude ratio of $\mathcal{P}_{\Pi}/\mathcal{P}_{\eta} \sim 10^{-3}$. The CMB anisotropies are significantly suppressed on small scales relative to large scales, and are similar to the scalar isocurvature mode. On small scales, the dark energy density is negligible at horizon crossing and the resulting growth in $\sigma$ is small. On larger scales, the dark energy density increases at horizon crossing resulting in larger growth in $\sigma$.

\subsection{Tensor Sector} \label{sec:tensobser}

\begin{figure}[] 
\centering
\mbox{\resizebox{0.4\textwidth}{!}{\includegraphics{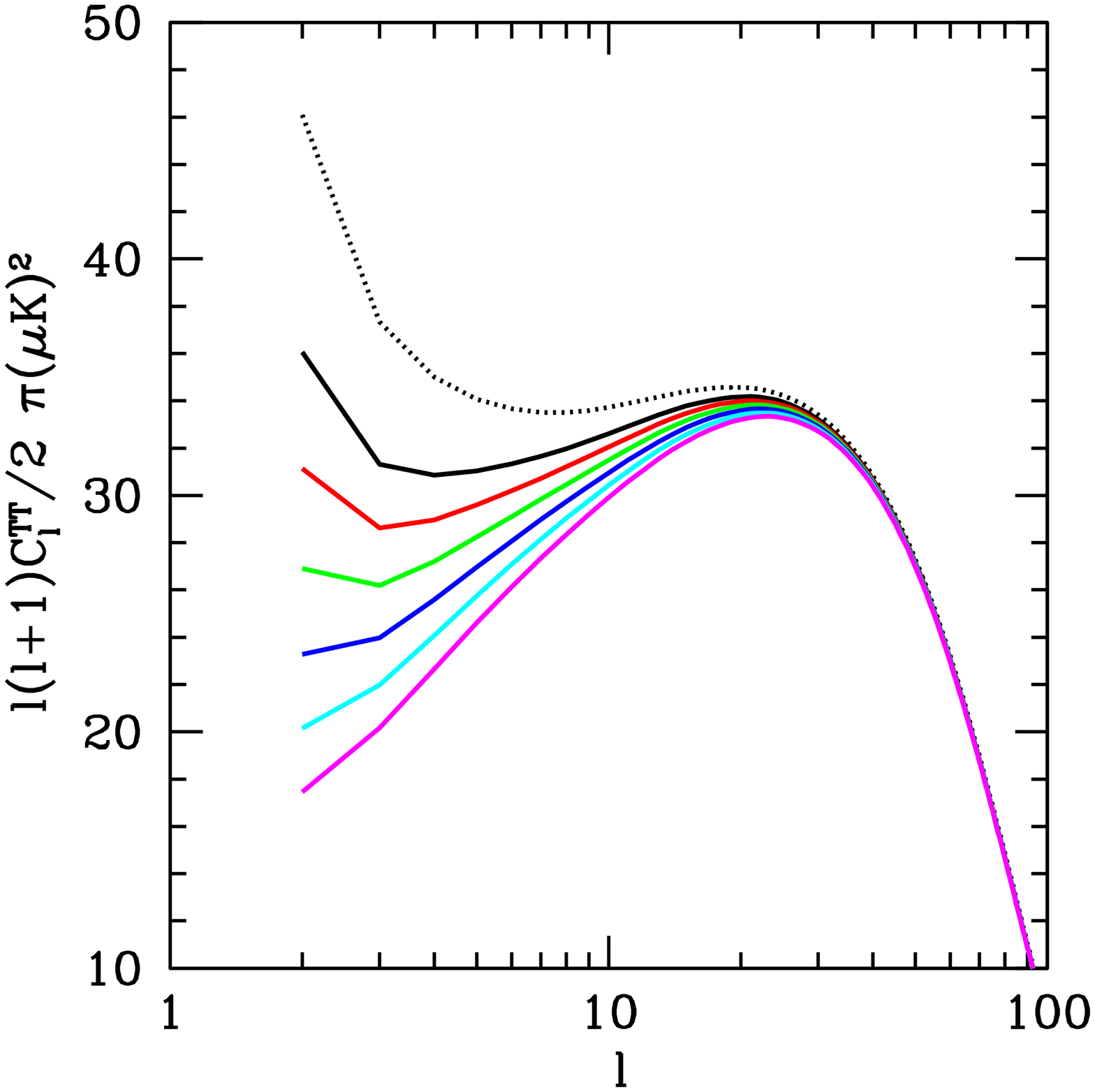}}}
\mbox{\resizebox{0.4\textwidth}{!}{\includegraphics{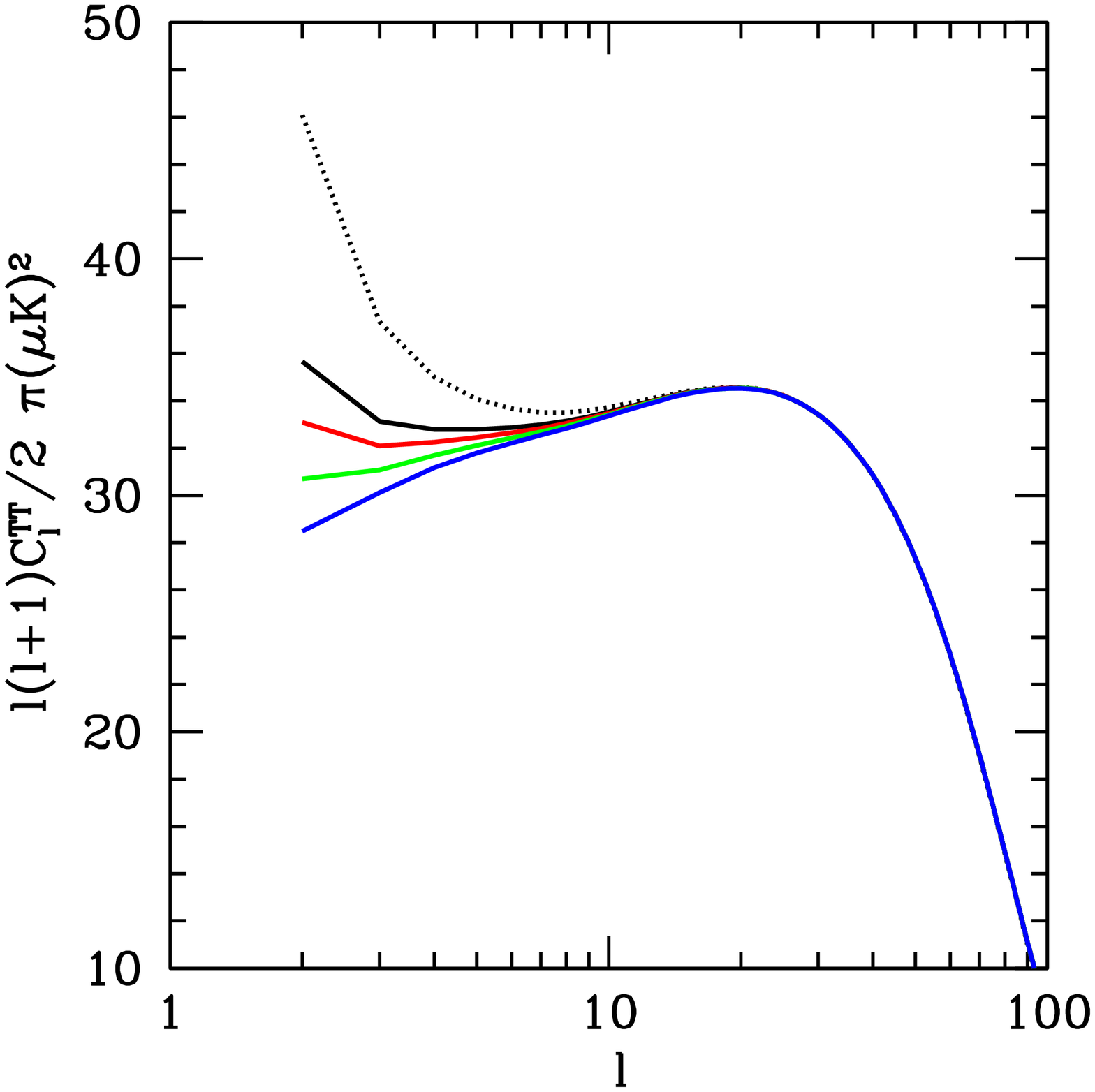}}}
\caption{\label{fig:tenstt} Tensor mode CMB TT power spectrum with $r=0.1$ at the pivot scale $k_{0}=0.05{\rm Mpc}^{-1}$.  The left panel shows the spectrum for a $w=-1/3$ elastic fluid and the right panel a $w=-2/3$ model.  The labelling of the curves is the same as in Fig.~\ref{fig:lambdacdmcmbTE}, and we observe a decrease in large scale power with increasing $c_{s}^{2}$.}
\end{figure}

In many inflationary models a nearly scale invariant spectrum of gravitational waves (tensor modes) is produced. These models are parameterized by the tensor spectral index $n_{t}$ and the tensor to scalar ratio $r=A_{t}/A_{s}$, where $A_{t}$ and $A_{s}$ are the primordial amplitude of tensor and scalar fluctuations and the tensor power spectrum is given by $\mathcal{P}_{t} (k) = A_{t} k^{n_{t}}$. The dominant tensor source to the CMB TT anisotropy is 
\begin{equation}
C_{\ell}^{TT} = 4 \pi \int d (\log k) \mathcal{P}_{t}(k) \left[ \int_{\tau_{\rm dec}}^{\tau_{\rm 0}} d \tau e^{- \kappa}  \dot{H}^{T} \frac{j_{\ell}(k(\tau_{0}-\tau))}{(k(\tau_{0}-\tau))^{2}} \right]^{2} \frac{(\ell+2)!}{(\ell-2)!}\, .
\end{equation}
For a scale invariant spectrum as $k \rightarrow 0$ the power in the photon Boltzmann hierarchy remains in the quadrupole and thus the tensor contributions have power enhanced on large angular scales. This is shown in Fig.~\ref{fig:tenstt}, where we plot the TT anisotropy for the $\Lambda$CDM model with $n_{t}=0$ and $r=0.1$ at the pivot scale $k_{0}=0.05{\rm Mpc}^{-1}$. The elastic fluid contributes a source of anisotropic stress to the Einstein equation~(\ref{eqn:evolution4}), and has the effect of damping the evolution of $H^{T}$ and reducing the power on large angular scales. This is another potential signature of elastic dark energy models.

\section{Possible realizations of elastic dark energy models}
\label{sec:micro}

One possible realization of the elastic fluid dark energy models is a frustrated network of non-Abelian cosmic strings~\cite{Vilenkin:1984rt,Kibble:1985tf,Spergel:1996ai,McGraw:1997nx} or domain walls~\cite{Battye:1999eq,Friedland:2002qs}. In this section we discuss some basic aspects of these scenarios and make attempts to link them with the earlier sections. In order to act as dark energy component, any model needs to provide significant negative pressure and have a Jeans length which is comparable to the present horizon, which can be achieved by having a relativistic sound speed.

The idea is that a static lattice of topological defects forms at some point after a phase transition in the Early Universe (see ref.~\cite{vilenkin:1994s} for a review of the physics of topological defects and phase transitions). Such a lattice will have microscopic, mesoscopic and macroscopic scales. The microscopic and mesoscopic scales refer to the core width and the lattice cell size respectively; we will not discuss these specifically here and will just assume that the relevant theory allows for the formation of a lattice which is stable to physics on these scales. Most important to the present discussion is the existence of some macroscopic scale, $L$.

On dimensional grounds the density of static cosmic strings scales as $\rho_{\rm str}\propto L^{-2}$ and that for domain walls is given by $\rho_{\rm dw}\propto L^{-1}$. The standard assumption of defect evolution based on simulations of the simplest field theoretic models is that a self-similar scaling regime is achieved (see, for example, ref.~\cite{Battye:2006pf} and references therein) whereby $L\propto t$. However, if a static, stable lattice forms, then the appropriate macroscopic length scale will scale with the expansion of the universe and $L\propto a(t)$. In this case the appropriate values of $w$ are $w=-1/3$ for strings  and $w=-2/3$ for walls and these have been the values which we have focused on in our earlier discussions. It is possible that other values of $w$ might be possible if the equation of state of the walls is not Nambu-Goto. For example,  the energy-momentum tensor of a string with energy density per unit length $U$ and tension $T$ is 
\begin{equation}
T_{\mu\nu}(x)=\int d\sigma\,\delta[x-X(\sigma)]\,\left(U\epsilon \dot X_{\mu}\dot X_{\nu}-{T\over \epsilon}X^{\prime}_{\mu}X^{\prime}_{\nu}\right)\,,
\end{equation}
where $\epsilon^2={\bf X}^{\prime\,2}/(1-\dot{\bf X}^2)$ and $X_{\mu}=(t,{\bf X}(\sigma))$ is the position of the string. Using this one can deduce that 
\begin{equation}
P_{\rm str}={1\over 3}\rho_{\rm str}\left[\left(1+{T\over U}\right)\langle v^2\rangle - {T\over U}\right]\,,
\end{equation}
where $\langle v^2\rangle^{1/2}$ is the rms velocity of the strings. Hence, one finds that $w=-T/(3U)$ in the static limit. Since causality implies that $T/U\le1$ we have that $w\ge -1/3$, with $w=0$ in the non-relativistic limit $T<<U$. Similar arguments can be put together to show that $w\ge -2/3$ for walls.

The static defect lattice will behave like an elastic solid since it has rigidity. Recently, it has been shown~\cite{Battye:2005hw} that under the assumption that the lattice is isotropic $\mu/\rho=4/15$ for static Dirac-Nambu-Goto strings and walls. If the rigidity is comparable to this then the lattice will be stable to macroscopic perturbation modes since $c_{\rm s}^2$ and $c_{\rm v}^{2}$ are both positive and the Jeans length will be a substantial fraction of the horizon since $c_{\rm s}, c_{\rm v} \sim 0.1$. It is clear that an exactly isotropic network would be difficult to form in cosmological phase transition since it will be uncorrelated on large scales. However, it might be possible for it to be approximately isotropic allowing for the treatment focused on in this paper to be applicable. An alternative is that the lattice has approximate point symmetry so the elasticity tensor will also have point symmetry as described in section~\ref{sec:aniso}. The shear moduli for the Bravais lattices with cubic symmetry (the primitive lattices: simple cubic (SC), face-centred cubic (FCC) and body-centred cubic (BCC)) relevant to the domain wall case were discussed in ref.~\cite{Battye:2005ik} and it was shown that if $w=-2/3$ then the BCC lattice is unstable and the SC/FCC lattices have zero modes. It was argued that for the SC lattice this mode corresponds to perturbation of infinite extent, that is, one of the faces of the cubic system moving toward the another. A number of composite lattices with cubic symmetry which are part of the tetrahedral close-packing (TCP) structures (see, for example ref.~\cite{kraynick:tcp}) are also possible, although none of these appear to be stable if $w=-2/3$. They could, however, be stable if $w=-2/3+\epsilon$ for $\epsilon$ small and positive. The values of shear moduli are tabulated in table~\ref{tab:shearvalues} for known cubic lattices.
 
\renewcommand{\arraystretch}{1.6}
\begin{table}
\begin{center}
\begin{tabular}{|c|c|c|c|}
\hline
   & $\mu_{\rm L}/\rho$ & $\mu_{\rm T}/\rho$ & $\Delta\mu/\rho$ \\
\hline
SC  & 1/6 & 1/3 & 1/6 \\
FCC & 2/9 & 1/6 & -1/18 \\
BCC & 0.109 & 0.183 & 0.074 \\
 Weaire-Phelan (A15) & 0.168 & 0.161 & -0.007 \\
 Bergmann (T) & 0.161 & 0.161  & $<10^{-3}$ \\
 Frauf-Laves (C15) & 0.159 & 0.166  & 0.007 \\
\hline
\end{tabular}
\end{center}
\caption{Shear moduli values for the cubic symmetry group with different primitive cells. $\Delta \mu$ is defined as $\mu_{\rm T}-\mu_{\rm L}$, and is zero in the isotropic case. The last three cases correspond to tetrahedral close-packing (TCP) structures (see, for example ref.~\cite{kraynick:tcp}).} 
\label{tab:shearvalues}
\end{table}

In addition to not being totally isotropic at formation (or even having {\it exact} point symmetry) a lattice may have initial macroscopic fluctuations relative to the equilibrium position which correspond to the isocurvature modes discussed in the previous section. Since the formation of the lattice is a causal process then these modes with have an initial white noise spectrum (${\cal P}_i(k)\propto k^3$).

Two interesting issues are the scale of symmetry breaking of the phase transition and the likely cell size of the lattice at the present day. Both of these rely on us understanding the lattice formation process which is not well-understood. A simple assumption would be that the lattice forms instantaneously after the phase transition with some initial cell size $\xi_{\rm c}=at_{\rm f}$ which is some fixed fraction of the horizon size at the time of formation $t_{\rm f}$. If $\eta$ is the symmetry breaking scale which relates to the temperature of formation $T_{\rm f}=b\eta$, then assuming the lattice forms in the radiation era the correlation size at formation is given by
\begin{equation}
\xi_{\rm c}(t_{\rm f})= \frac{0.3a}{b^{2}} \mathcal{N}^{-1/2} \frac{m_{\rm pl}}{\eta^{2}}\,,
\end{equation}
where $\mathcal{N}$ is the number of relativistic species at formation. In the case of domain walls, if the wall density at formation is $\rho(t_{\rm f})=c \eta^{3}/\xi_{\rm c}$, and assuming the network is subsequently swept along by the Hubble flow, we can estimate $\eta$ to be
\begin{equation} \label{eqn:etawall}
\eta=100 {\,\rm keV} \left( \frac{a}{c \,b}\right)^{1/4} \left( \frac{\mathcal{N}}{100} \right)^{-1/8} \left(\Omega_{\rm dw} h^{2}\right)^{1/4}\,.
\end{equation} 
Since the coefficients representing uncertainty in the wall formation process have small exponents, it seems reasonable to estimate $\eta\approx 100\,$keV if $\Omega_{\rm dw} h^{2} \sim 1$. One can also compute the present day cell size in a similar fashion, which turns out to be
\begin{equation}
\xi_{\rm c}(t_{0})=100 {\,\rm pc} \left( \frac{a^{3} \,c}{b^{3}}\right)^{1/4} \left( \frac{\mathcal{N}}{100} \right)^{-3/8} \left(\Omega_{\rm dw} h^{2}\right)^{-1/4}\,.
\end{equation}
Since the smallest wavenumber at which non-linear effects become important in CMB codes such as CAMB is $k\sim0.2\, {\rm Mpc^{-1}}$, it seems reasonable to assume that the linear response to perturbations in such a defect network can be treated in the continuum elastic medium framework. The computation of the cell size is more sensitive to the uncertainty parameters, but since causality gives the upper limit of $a \lesssim 1$ the continuum medium description seems justified. One can also perform a similar exercise in the case of cosmic strings. If the wall density at formation is $\rho(t_{\rm f})=c \eta^{2}/\xi_{\rm c}^{2}$ then one finds $\eta\approx 2\,$TeV and $\xi_{\rm c}(t_{0})\approx1.0\,$AU, with similar uncertainties in these parameters as in the domain wall case.

The allowed symmetry breaking scale may be increased further if the defect network forms during inflation. In this scenario the network is inflated outside the horizon and re-enters during a later epoch thus diluting the initial density. If one assumes that $e^{N}=T_{\rm R}/T_{\rm f}$ where $T_{\rm R}$ is the reheat temperature of inflaton and $N$ the number of e-folds remaining when the network leaves the horizon during inflation then the symmetry breaking scale of a domain wall network is given by
\begin{equation}
\eta= {5\times10^{-7}\,\rm GeV} \left( \frac{a}{c}\right)^{1/3} \left( \frac{\mathcal{N}}{100} \right)^{-1/6} \left( \frac{T_{\rm R}}{1 \,\rm TeV}\right)^{-1/3} e^{N/3}\left(\Omega_{\rm dw} h^{2}\right)^{1/3} \,.
\end{equation}
and the present day cell size is modified to 
\begin{equation}
\xi_{\rm c}(t_{0})=10^{-5} {\,\rm pc} \,\,a \left( \frac{\mathcal{N}}{100} \right)^{-1/2} e^{N} \left( \frac{1\,\rm TeV}{T_{\rm R}} \right)\,.
\end{equation}
If $N=30$ and $T_{\rm R}=1\,$TeV then $\eta\approx10\,$MeV and $\xi_{\rm c}(t_{0})\approx100\,$Mpc. One should note that quantum fluctuations in the defect forming field, $\psi$, should satisfy satisfy $\delta \psi \sim H <\eta$, otherwise no phase transition can occur during inflation. If $V \sim M^{4}$ this restricts the mass scale of inflation to $M<3\times 10^{8}\,$GeV.

If a lattice of domain walls or cosmic strings exists, their local gravitational interactions can give rise to further observable effects. The symmetry breaking scale $\eta$ is limited observationally by the local generation of density induced perturbations. In the case of domain walls, the mass per unit area is $\sigma\sim\eta^{3}$ and so $\eta=100\,$keV gives $\sigma\sim5\times10^{-8}\,\rm{kg\,m}^{-2}$, and for $\eta=10\,$MeV the wall has $\sigma\sim5\times10^{-2}\,\rm{kg\,m}^{-2}$. If there are only several walls in our current horizon they induce a density fluctuation $\delta \rho/\rho\sim G\sigma t_{0}$, with a similar sized fluctuations in temperature of the CMB. A wall with $\eta=100\,$keV would induce direct temperature fluctuations of $\delta T/T\sim10^{-9}$, which is well below the observed value of $\delta T/T\sim10^{-5}$. Using this constraint directly, one find that domain walls with $\eta\gtrsim 1\,$MeV would be ruled out~\cite{Zeldovich:1974uw}. However, this assumption assumes that the walls move relativistically, with only several walls in our horizon. In the lattice structures we have envisaged here the walls are static and only move with Hubble flow, and the CMB distortion in this case can be much smaller~\cite{nambu:1991cmb}. In the case of strings, the mass per unit length is given by $\mu\sim\eta^{2}$ and so $\eta=1\,$TeV gives $\mu \sim9\times10^{-6}\,\rm{kg\,m}^{-1}$. The CMB temperature distortion  is of the order $\delta T/T \sim 8 \pi G \mu$, and so low energy strings discussed here would not produce significant fluctuations in temperature.

\section{Discussion and Conclusions} \label{conclusion}

We have studied the cosmological implications of a perfect elastic fluid in the framework of General Relativity. In previous work on this subject, Bucher and Spergel derived the equations of motion by variation of the action assuming that the fluid is isotropic~\cite{Bucher:1998mh}. Here, we take a more general approach using the material representation concept and obtain equations of motion in terms of the pressure and elasticity tensors of the fluid. This allows us to parameterize the most general description of linearized perturbations of the fluid in a compact and transparent manner.

Under the assumption that the pressure and elasticity tensors are isotropic we derive the Einstein and energy-momentum conservation equations using this approach. When the fluid has non-zero rigidity there is a source of anisotropic stress which stabilizes perturbations when $w<0$, making these models candidates to describe the dark energy. The anisotropic stress also interacts with the vector and tensor sectors as the shear modulus induces transverse waves in the fluid. This phenomenon is well known in the laboratory. In a non-relativistic (low pressure) deformed medium with non-zero rigidity the temperature has both temporal and spatial variations. If the transfer of heat occurs slowly then the oscillatory motions in the deformed body are adiabatic. These motions correspond to longitudinal (scalar) and transverse (vector) elastic waves, which are related by $c_{\rm s}^{2} > 4 c_{\rm v}^{2}/3$~\cite{landau:1959}. In the relativistic (high pressure) treatment we enforce adiabicity by assuming that there are no temporal or spatial variations in $w$ and the relationship between the wave propagation modes is then given by $c_{\rm s}^{2}=w+4 c_{\rm v}^{2}/3$.

We find that the elastic fluid model is similar to generalized dark energy models~\cite{Hu:1998tj, Hu:1998kj}. The construction of the fluid energy-momentum tensor in these models was phenomenological, and it was argued that anisotropic stress can be attributed to viscosity in the fluid. We find that the functional form of anisotropic stress suggested in~\cite{Hu:1998tj, Hu:1998kj} is identical to our own if the decay timescale associated with the stress term is infinite. In this case, the viscous sound speed can be directly related to the vector sound speed of a fluid with rigidity by $c_{\rm vis}^{2}=c_{\rm v}^{2}(1+w)/2.$ 

There appears to be two mechanisms for stabilizing dark energy perturbations if we wish to model the dark energy as a fluid. Either the fluid can be adiabatic and have rigidity, or it can be non-adiabatic and have entropy perturbations. A microphysical realization of the latter possibility is scalar field dark energy. As the evolution of perturbations is different in each model, we have investigated the potential observational differences in the CMB and matter power spectrum assuming the initial conditions were generated from curvature fluctuations in the metric. We find that the possibilities for distinguishing the two models are low using only CMB TT data. However, the TE power at large scales is reduced in the elastic fluid models relative to the $\Lambda$CDM and scalar field models with $\kappa_{\rm R}\ne0$. This increases the parameter space in which the two models can be differentiated, up to $w \gtrsim -0.8$. As $w$ becomes closer to -1 the dark energy perturbations become less important, and the two models essentially become the same from an observational point of view.
 
The matter power spectrum outlines an interesting difference between the two models. An important aspect of the dark energy is that it has a large Jeans length so that it does not cluster on small scales. In principle though, it can cluster on large scales and we have attempted to model this by including dark energy perturbations in $P_{\delta}(k)$. As $c_{\rm s}\rightarrow 0$ we find that the elastic dark energy contribution to $P_{\delta}(k)$ dominates and $c_{\rm s}\gtrsim 10^{-3}$ would be required to be compatible with large scale structure data. In the case of scalar field dark energy the contribution to $P_{\delta}(k)$ from dark energy increases as $c_{\rm s}\rightarrow 0$ but remains sub-dominant even when $c_{\rm s}^{2}=0$. 

We have also shown that both scalar and vector isocurvature modes are allowed in an elastic fluid. These modes correspond to initial fluctuations of the fluid relative to the background, and give rise to anisotropic stress on super-horizon scales. A generic feature of these modes is that power is suppressed on small relative to large scales. The dominant mode in these models would have to be the adiabatic curvature mode in order to fit the data but there could be a sub-dominant isocurvature component.

Our approach has shown that we can parameterize the perturbations of the energy momentum tensor in terms of the pressure and elasticity tensors of the fluid. In this work we mainly focused on the isotropic case, but this formalism can be easily extended if the elastic fluid has point symmetry. The symmetry properties of the elasticity tensor can be classified by the Bravais lattices, which are familiar in crystallography. Recently, there has been evidence of large scale anomalies in the CMB, most noticeably the North-South power asymmetry~\cite{Eriksen:2003db} and alignment of low $\ell$ multipoles~\cite{Tegmark:2003ve,deOliveira-Costa:2003pu,Schwarz:2004gk,Land:2005ad}, which appear to persist in the third year WMAP data~\cite{Copi:2006tu}. A possible solution is that the energy momentum tensor of the dark energy is not rotationally invariant at linearized order. We have considered the case where the elasticity tensor has cubic symmetry and find that perturbation growth is dependent on the direction as well as the magnitude of the wave vector. This is work is developed further in ref.~\cite{Battye:2006mb}, and we are currently investigating whether these anisotropic models are compatable with the data.

\begin{acknowledgments}

We would like to thank Martin Bucher, Brandon Carter and Elie Chachoua for useful discussions. 

\end{acknowledgments}

\end{document}